\documentstyle[aps,twocolumn,floats,prl]{revtex}
\input epsf.sty

\newcommand{\vertsp}{\vphantom{\dot a \over a}}
\newcommand{\stress}{{S}}
\newcommand{\stressa}{{S_\Pi}}
\newcommand{\stressad}{{S_A}}
\newcommand{\stresssonic}{S_{S}}
\newcommand{\stressna}{S_{\Gamma}}
\newcommand{\stressg}{S_{E}}
\newcommand{\potential}{{A}}
\newcommand{\shift}{{B}}
\newcommand{\curvature}{{H}_L}
\newcommand{\shear}{{H}_T}
\newcommand{\entropy}{\sigma}
\newcommand{\rhocr}{\rho_{\rm cr}}
\newcommand{\rhoseed}{\rho_{\rm s}}
\newcommand{\pseed}{p_{\rm s}}
\newcommand{\vseed}{v_{\rm s}}
\newcommand{\Piseed}{\Pi_{\rm s}}
\newcommand{\zetac}{{\zeta}}
\newcommand{\xic}{{\xi}}
\newcommand{\rhoc}{{\rho}}
\newcommand{\rhoC}{\rho_C}
\newcommand{\pc}{{p}}
\newcommand{\pC}{{p_C}}
\newcommand{\Pic}{{\Pi}}
\newcommand{\vc}{{v}}
\newcommand{\vC}{{v_C}}
\newcommand{\wc}{{w}}
\newcommand{\wC}{{w_C}}
\newcommand{\wcr}{w_{\rm cr}}

\begin{document}

\twocolumn[\hsize\textwidth\columnwidth\hsize\csname
@twocolumnfalse\endcsname

\title{The Structure of Structure Formation Theories}

\author{Wayne Hu and Daniel J.\ Eisenstein}
\address{Institute for Advanced Study, Princeton, NJ 08540}
\maketitle
\begin{abstract}
We study the general structure of models for structure formation,
with applications to the reverse engineering of the model
from observations.
Through a careful accounting of the degrees of freedom in
covariant gravitational instability theory,
we show that the evolution of structure is completely 
specified by the stress history of the dark sector.  
The study of smooth, entropic,
sonic, scalar anisotropic, vector anisotropic, and tensor
anisotropic stresses reveals the origin, robustness, and uniqueness
of specific model phenomenology.
We construct useful and illustrative analytic solutions 
that cover cases with multiple species of differing
equations of state relevant to the current generation of
models, especially those with effectively smooth components.  
We present a simple case study of models with phenomenologies
similar to that of a $\Lambda$CDM model to highlight 
reverse-engineering issues.  A critical-density universe dominated
by a single type of dark matter with the appropriate stress
history can mimic a $\Lambda$CDM model exactly.
\vskip 1truecm

\end{abstract}
]

\section{Introduction}

How does one reverse engineer a model for structure formation from
observed phenomena?  How unique is such an inversion?
How robust are the phenomenological distinctions between
broad classes of models?
With the wealth of high-precision cosmological data 
expected in the near future from the cosmic microwave
background (CMB), galaxy surveys, and the high redshift universe,
the simple {\it ab initio} models for structure formation
currently considered may be ruled out, forcing us to 
confront these difficult issues.  In this paper, we take the
first steps toward answering this question by examining from
a general standpoint what makes a model for structure formation
behave as it does in linear perturbation theory.

A model for structure formation 
is completely specified by its initial conditions and the 
full temporal and spatial behavior of the stresses in
its dark sector.  
The dark sector contains the elements in the model that
do not interact with the photons at any observable redshift.
It can include, but is not limited to, cold dark matter (CDM),
neutrinos, and cosmological defects.

Unfortunately, the stress history
of the dark sector is by definition not directly observable.
Its effects come filtered through gravity as mediated
by metric fluctuations.  The translation of metric fluctuations
into observables in the CMB and evolution of structure is
well understood.  Therefore, the main hurdle in the
task of reconstructing a model from observations
is to understand how stress histories
translate into metric fluctuations and vice versa.

Our general philosophy here is to start from elements of
the cosmological model that will likely survive the onslaught of
data: general relativity and a universe whose deviations from
homogeneity and isotropy are initially small.  We
proceed down the theory pipeline to the existing models
of structure formation, making explicit 
the places where assumptions are made and hence could 
be altered.  Where possible, we provide analytic solutions
and approximations that highlight certain generic behavior 
and phenomena.   These solutions are useful for
describing the behavior of the existing models and 
are in most cases new or substantially more general than
those found in the literature.  In particular, we derive
master solutions for models which contain multiple 
components with 
arbitrary 
equations of state and smooth, entropic, anisotropic,
and sonic stresses.

We begin in \S \ref{sec:taxonomy} 
with an overview of the basic elements of a structure
formation theory and their traditional classification 
in terms of their 
initial conditions, perturbation type, and clustering
properties.  We then present a concise
but general treatment of linear perturbation theory in \S 
\ref{sec:perturbationtheory} and gauge issues in \S
\ref{sec:gauge}.  Although these are well-studied
subjects (see e.g. \cite{Bar80,KodSas84,MukFelBra92}), 
our treatment has several pedagogical and practical 
virtues. It keeps careful track of the degrees of freedom available 
to structure formation models and hence provides a unified
treatment applicable to all models, including those containing
exotic matter like scalar fields or cosmological defects.  
We also explicitly maintain general covariance
such that the equations apply, and may be easily 
specialized, to any choice of coordinates or gauge. 
In \S \ref{sec:stressrepresentation}, we define
general classes of stress perturbations and present an 
overview of their conversion into observables. 

The remainder of the paper deals with stress histories on
a case by case basis.  The simplest case involves
smooth stresses, and we present
detailed analytic solutions in \S \ref{sec:stressfree}
that apply to a wide range of models---from 
simple cosmological constant models ($\Lambda$CDM)
to massive neutrino
and scalar field models. 
Pure anisotropic, entropic, and sonic stresses are treated in
\S \ref{sec:purestress} and mixed cases in \S \ref{sec:mixedstress}.
To highlight reverse-engineering issues, we study
single-component, critical-density dark matter models with 
phenomenologies that mimic the $\Lambda$CDM model 
in \S \ref{sec:designer}.
We conclude in \S \ref{sec:conclusions} by re-examining the
traditional classification scheme of \S \ref{sec:taxonomy} in
light of the phenomenological distinctions uncovered in this
work.

\section{Classification of Theories}
\label{sec:taxonomy}
\subsection{Initial Conditions}

Perhaps the most fundamental difference between models for structure
formation lies with their initial conditions.  Currently, inflation is
the only known means of laying down large-scale density or curvature
perturbations in the early universe.  Indeed, inflation in the
more general sense of a period of superluminal expansion is 
required for the causal generation of large-scale power.   
It provides a means of connecting parts of the universe that are
currently space-like separated, i.e.~outside the current particle
horizon.   Models with initial curvature perturbations are usually 
called ``adiabatic'' models.

All other causal models begin with no density or curvature 
fluctuations on large scales and are hence called ``isocurvature''
models.  
In these models, stress gradients causally move matter
around inside the horizon to form large-scale structure. 

The generation mechanism is also responsible for determining
the spectrum and statistics of the fluctuations.  
The simplest inflationary
models predict a nearly scale-invariant and 
gaussian distribution of fluctuations \cite{BarSteTur83}
but higher order effects can break scale-invariance and
generate non-gaussianity
\cite{AllGriWis87,MosMatLucMes91}.
Defect perturbations are intrinsically non-gaussian but are typically also
scale-invariant in the generalized sense of ``scaling" 
\cite{Kib85}.
We are primarily concerned here with the evolution 
of fluctuations from their initial state through the linear
regime and do not consider these issues further. 
Note that changes in the
spectrum of perturbations are simple to 
include in linear theory as evolutionary effects can 
be factored out into so-called ``transfer functions".

\subsection{Perturbation Type}

The
perturbation type for the metric and matter fluctuations is the next most important distinction.  
A general linear fluctuation can be decomposed 
into scalar, vector and tensor components.  These manifest themselves
as density, vorticity, and gravitational wave perturbations 
respectively and do not interact in linear theory.  The scalar
modes are the only ones that grow through gravitational 
instability.  
Vector modes, on the other hand, always decay with the expansion.  They can only
be actively generated
by shearing (or anisotropic) stress in the manner.
Tensor modes are intermediate.  Left to themselves, they propagate
as gravity waves, but they generate and can be generated by
transverse-traceless (quadrupolar) stresses in the matter. 

The simplest inflationary models possess only scalar (``S'') 
fluctuations, 
and tensor fluctuations are generally cosmologically negligible in models with
energy scales substantially below the Planck scale \cite{Lyt97}.  
However, models whose initial conditions contain both
scalar and tensor (``ST'') 
fluctuations are possible.  Models with only S or ST generally
have stresses that may be defined as functions of the
metric, density and velocity perturbations and hence may be
viewed as ``passive'' responses through equations of state.  

``Active'' models have stresses that are a consequence of
complex internal dynamics in the dark sector that cannot 
be simply specified as responses to gravitational perturbations.
Although this definition is not precise for scalar and tensor 
modes, the very presence of vector modes indicates an active
source because these must be continuously generated to have
an observable effect. 
Nevertheless, 
such models generally have all three types of perturbations (``SVT'').

\subsection{Clustering Properties of Dark Matter}

Finally, the nature of the dark components affects the evolution of
perturbations.  We define as ``dark" any component that interacts
with the CMB photons only gravitationally.  Thus, even massless
neutrinos are classified as dark matter in this scheme.

Stresses in the dark components change the evolution of
the mean density with time and the response of the matter to
gravitational compression.  We will loosely type models whose
expansion rate is driven by a compressible type of matter 
(on scales relevant to cosmological structures)
as
``clustered'' models and those which possess
matter that is incompressible 
 as ``smooth'' models.  We shall
see that this distinction is in fact rather inexact as it is
not time invariant: essentially all models pass through phases when 
they would be considered smooth or clustered on the relevant
scale.  

\subsection{Phenomenology}
\label{sec:generalphenomenology}

\begin{figure}[tb]
\centerline{ \epsfxsize = 3.5truein \epsffile{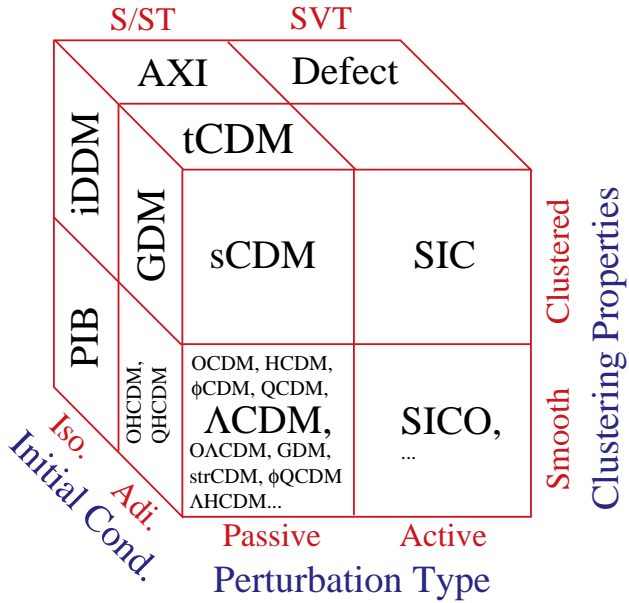}}
\caption{Taxonomy of structure formation.  
Models can be classified by their
initial conditions (adiabatic or isocurvature), perturbation type
(passive or active), and clustering properties of the
dark matter (clustered or smooth on large-scale structure scales).
Passive fluctuations involve stress responses to other perturbations
and can support scalar (``S'') or scalar and tensor
(``ST'') components. 
Active stresses generate fluctuations and
generally possess vector components as well (``SVT''). 
We will examine the extent to which this traditional categorization
is useful in predicting model phenomenology.}
\label{fig:taxonomy}
\end{figure}

The key to understanding the phenomenology of a given model for 
structure formation is the evolution of metric fluctuations, in 
particular the Newtonian gravitational potential. 
Its qualitative behavior is determined by the initial conditions,
perturbation type, and dark matter content 
of the model. We illustrate this taxonomy scheme in
Fig.~\ref{fig:taxonomy}.

The behavior of the gravitational potential is directly related 
to the evolution of density perturbations through the 
Poisson equation.  Once its evolution is determined as a function
of scale, not only is the present large-scale structure of the
universe determined but also the whole time history of
structure formation.
The latter is important for predicting the properties and abundances
of high-redshift objects.

The gravitational potential also generates CMB anisotropies through
gravitational redshifts \cite{SacWol67}
and is the ultimate source of all anisotropies
from scalar perturbations. CMB phenomenology 
can be essentially read off of 
the time evolution of the gravitational potential \cite{HuSug95},
although this involves understanding the backreaction from density
perturbations in the CMB itself \cite{HuWhi96}.   

Similar but simpler considerations apply for vector and tensor
metric perturbations.  They also generate anisotropies via 
gravitational redshifts but do not have unstable modes 
and hence do not affect large-scale structure formation
in linear theory.  

The difference between adiabatic
and isocurvature models 
play a direct role in metric evolution because 
initial curvature (or gravitational potential)
perturbations are present in one and absent in the other.
The 
perturbation type changes the ratio of  CMB anisotropies
to large-scale structure.
Finally, the dark
matter properties affect the evolution of the gravitational potentials.
Smooth components by definition do not contribute to the gravitational
potential but do contribute to the expansion rate.  They slow
down the growth of structure and cause the gravitational potential
to decay.  Hence they decrease the amount of structure and increase
the large-angle anisotropies of the CMB.  

In summary, the observable properties of structure formation models
are encapsulated in the time evolution of the metric fluctuations.
This in turn is governed by the stress properties of the matter
both through its initial conditions and intrinsic properties.

\subsection{Current Model Zoo}
\label{sec:currentzoo}

The archetypal model for structure formation is the standard
cold dark matter model (sCDM), which is an adiabatic, passive,
and clustered model.
Here, an initial scale-invariant 
spectrum of adiabatic scalar (``S'') perturbations
collapses via the gravitational instability of pressureless
cold dark matter.  Although this model is no longer viable from
an observational standpoint, it predicts
phenomena sufficiently similar to the observations 
to act as a good starting 
point for model building.  
One of its failings is that it predicts too much 
small-scale power for the level of
CMB anisotropies demanded by the COBE detection.
 
A simple variation of the sCDM model that attempts to address this
problem involves tilting the initial
spectrum of scalar perturbations (tCDM)
to reduce small-scale power relative
to large.  Under certain inflationary
scenarios, this brings about the addition of tensor perturbations
that further reduce small-scale density perturbations
relative to the COBE detection.  
Such a model would be a ``ST'' variant 
of sCDM. 

The second class of variations involves changing the matter
content so as to suppress the clustering of matter, 
yielding a ``smooth'' variant.  
The prototypical example is the $\Lambda$CDM model, where 
an additional component of matter that does not cluster replaces
most of the CDM.  Another examples is the OCDM model where
spatial curvature plays the role of the smooth component. 
Those two represent examples where the additional component
is smooth on all scales and for all time by definition.
Variants where the matter is only smooth on small scales
include the HCDM model (e.g. \cite{HolPri93} also called
C$+$HDM and MDM) 
with a component of
hot dark matter, the $\phi$CDM  model \cite{FerJoy97}
with a scalar
field component $\phi$ that tracks the background 
behavior of the matter, and  QCDM \cite{QChicago,QPaul} 
with a general scalar field
(``quintessence'').  
In a string-dominated universe (strCDM) 
\cite{SpePen97}, the string network plays the role of a smooth
component with the same equation of state as spatial curvature.  
Of course, one can have multiple smooth
species as well, e.g. O$\Lambda$CDM.
The GDM class of models \cite{Hu98}
phenomenologically
parameterizes all such models. 

Replacing the CDM with GDM of a different equation of state
but no stress perturbations is a phenomenological
possibility (GDM) suggested by \cite{Hu98}.  This is an
adiabatic, passive, and clustered variant of CDM.  
We will 
use the designation ``CDMv'' to represent all such variants of the
CDM model. 

Isocurvature models have been proposed as alternatives to the
CDMv class of models. The simplest examples are those
in which the initial stress fluctuations are established by
balancing the density perturbations of two different types of
matter.  Examples include the axion isocurvature (AXI) 
\cite{EfsBon86}
model and the primordial isocurvature baryon (PIB) model 
\cite{Pee87}, 
where radiation density
fluctuations are balanced by axions and baryons, respectively.
The simplest versions involve only scalar fluctuations and hence are
passive (``S'') models.  Models with and without smooth
cosmological constant or spatial curvature components
have been proposed.
Versions with gaussian power-law initial conditions are observationally
challenged \cite{EfsBon86,HuBunSug95} but more complicated variations
exist \cite{Pee97}.  Based on our work, one of us has constructed
an isocurvature decaying dark matter (iDDM) model that 
defies conventional wisdom on isocurvature models and solves
these observational problems \cite{Hu98b}

Finally, topological defect models such as strings and textures
fall into the isocurvature class but have fluctuations that
are active (``SVT").  The simplest versions obey 
scaling and have only clustering matter but fail
to generate enough large-scale structure for the observed
CMB anisotropies \cite{PenSelTur97,AlbBatRob97a}.
Models with a smooth $\Lambda$ component
have been proposed to alleviate these problems
\cite{AlbBatRob97b}.

Hybrid models can also be constructed.  If defects form
after the inflationary epoch, one has a model with adiabatic
initial conditions and active perturbations.  A string model
with inflation and cold dark matter (SIC) is a concrete
example \cite{ConHinMag98}.  One can also add in smooth
components, e.g. spatial curvature (SICO).

Clearly, the existing models do not even qualitatively exhaust
the possibilities open to structure formation models.  In the rest
of the paper, we conduct an examination of these 
possibilities beginning with general principles and explicitly 
stating the assumptions that are made in obtaining the models
described as well as their generalizations.  We summarize this
analysis in a series of flowcharts 
(Figs.~\ref{fig:einsteinsvt}, \ref{fig:scalar}, \ref{fig:vector},
\ref{fig:tensor}, \ref{fig:stressfree}, and \ref{fig:scalarstress}.) 

\section{Covariant Perturbation Theory}
\label{sec:perturbationtheory}

\subsection{General Definitions}

We assume that the background is described by an FRW metric
$\bar g_{\mu\nu} = a^2 \gamma_{\mu\nu}$
with scale factor $a(t)$ normalized to unity today
and constant comoving curvature
in the spatial metric $\gamma_{ij}$.
Here greek indices run from $0$ to $3$ while latin indices run over the
spatial part of the metric: $i,j=1,2,3$.
The component corresponding to conformal time
\begin{equation}
x^0 \equiv \eta = \int {dt\over a(t)}
\end{equation}
is $\gamma_{00}=-1$ and $\gamma_{0i}=\gamma_{i0}=0$. Unless
otherwise specified, overdots represent derivatives
with respect to conformal time and primes derivatives
with respect to $\ln a$.  $c=1$ throughout.
The background curvature is given by 
$K = - H_0^2(1-\Omega_{\rm tot})$, where the Hubble constant
is $H_0= 100 h\, $ km s$^{-1}$ Mpc$^{-1}$.  

The ten degrees of freedom for the perturbations in
the symmetric metric tensor $g_{\mu\nu}$ can be parameterized as
\begin{eqnarray}
g^{00} &=& -a^2(1-2 \potential)\,, \vertsp\nonumber\\
g^{0i} &=& -a^2 \shift^i\,,  \vertsp\nonumber\\
g^{ij} &=& a^2 (\gamma^{ij} -
        2 \curvature \gamma^{ij} - 2 \shear^{ij}) \vertsp\,,
\label{eqn:metric}
\end{eqnarray}
We refer to the lapse $\potential$ as the potential, the
three components of $\shift_i$
as the metric shift,
$\curvature$ as the curvature perturbation, and the
five components of $\shear^{ij}$ 
as the metric shear following the conventions of 
\cite{Bar80,KodSas84}.

Likewise, the symmetric stress-energy tensor can be parameterized
by ten components
\begin{eqnarray}
T^0_{\hphantom{0}0} &=& -\rho - \delta\rho\,, \nonumber\vertsp\\
T^0_{\hphantom{0}i} &=& (\rho + p)(v_i - \shift_i) \,, \nonumber\vertsp\\
T_0^{\hphantom{i}i} &=& -(\rho + p)v^i\,, \nonumber\vertsp\\
T^i_{\hphantom{i}j} &=& (p + \delta p) \delta^i_{\hphantom{i}j} 
	+ p\Pi^i_{\hphantom{i}j}\,, \vertsp
\label{eqn:stressenergy}
\end{eqnarray}
i.e. the energy density and its perturbation $(\rho+\delta\rho)$,
the isotropic stress (pressure)  
and its perturbation $(p + \delta p)$, the 
three components of the momentum density $(\rho+p)v_i$,
and the five components of the anisotropic stress tensor $\Pi_{ij}$. 
Note that
the metric shift $\shift_i$ enters in $T^0_{\hphantom{0}i}$
but not $T_0^{\hphantom{i}i}$.  Correspondingly, we shall
see that $B_i$ enters into the momentum but not the
energy conservation equation. 

By writing the metric and stress energy tensor in this form,
we have maintained general covariance.
As a result, the equations of motion that result below 
take the same {\it form} for any coordinate system where linear perturbation 
theory holds.
We reserve the term gauge invariant refers for objects that have 
the same {\it value} in each frame.

\begin{figure}[tb]
\centerline{ \epsfxsize = 3.4truein \epsffile{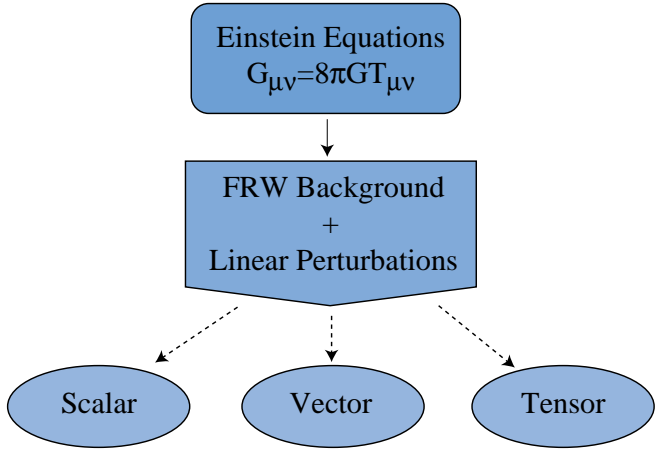}}
\vskip 0.5truecm
\caption{Scalar, vector, tensor decomposition.  At the top of the tree of
possibilities for structure formation models is the assumption that general
relativity holds in the cosmological context and universe is homogeneous
and isotropic in the mean with linear perturbations initially.  Without
further assumptions, the linear fluctuations may be expanded in scalar,
vector, and tensor modes that do not interact while the fluctuations remain 
linear.} 
\label{fig:einsteinsvt}
\end{figure}

\subsection{Perturbation Representation}

While the perturbations are linear, they may be separated
by their transformation properties under rotation without
loss of generality (see Fig.~\ref{fig:einsteinsvt}).  
For covariant techniques that do not assume linear perturbations
from the outset, see \cite{GebEll98,ChaLas98} and references 
therein.
The five component 
metric shear $\shear^{ij}$ and
matter anisotropic stess $\Pi^{ij}$ separate into one scalar,
two vector, and two tensor components.  The scalar stress/shear 
generates potential flows $({\boldmath \nabla \unboldmath}
\times {\bf v}=0)$, whereas the vector stress/shear generates
vorticity $({\boldmath \nabla} \cdot {\bf v}=0)$.  
The tensor stress generates tensor shear 
in the metric,
which represents gravity waves in the transverse-traceless gauge.  
Therefore, the potential $\potential$, curvature $\curvature$,
density $\delta\rho$, and pressure $\delta p$ perturbations
are associated with scalar fluctuations alone, the metric shift
$\shift_i$ and velocity $v_i$ with scalar and vector fluctuations,
and the metric shear $\shear^{ij}$ and anisotropic stress $\Pi^{ij}$
with all three.
Scalar, vector and tensor perturbations
may be treated independently in linear perturbation theory.

Fluctuations can be decomposed into the normal modes of 
the Laplacian operator \cite{Bar80}
\begin{equation}
\begin{array}{rcll}
\nabla^2 Q^{(0)} &= &  -k^2 Q^{(0)} \quad &{\rm S} \,,\vertsp \\
\nabla^2 \displaystyle{Q_i^{(\pm 1)}}
	   &= & -k^2 Q_i^{(\pm 1)} \quad &{\rm V}\vertsp \,, \\
\nabla^2 Q_{ij}^{(\pm 2)} 
	   &= &  -k^2 \displaystyle{Q_{ij}^{(\pm 2)}} \quad \vertsp &{\rm T} \,, \\
\end{array}
\end{equation}
where vector and tensor modes satisfy a divergenceless and
transverse-traceless condition respectively
\begin{eqnarray}
\nabla^i Q_i^{(\pm 1)} = 0\,, \quad
\gamma^{ij} Q_{ij}^{(\pm 2)} = \nabla^i Q_{ij}^{(\pm 2)} = 0 \,. 
\end{eqnarray}
In flat space, these correspond to plane waves times 
a local angular basis for the vectors and tensors
\cite{HuWhi97}.

Vector and tensor objects can of course be built out of scalar
and vector normal modes through covariant differentiation 
and the metric tensor \cite{KodSas84}
\begin{eqnarray}
Q_i^{(0)} & = &  -k^{-1} \nabla_i Q^{(0)}\,, \nonumber \\
Q_{ij}^{(0)} & = &  
	(k^{-2} \nabla_i \nabla_j - {1 \over 3} \gamma_{ij}) 
	Q^{(0)}\,,  \\
Q_{ij}^{(\pm 1)} & = &  -{1 \over 2k}[ \nabla_i Q_j^{(\pm 1)} 
	+ \nabla_j Q_i^{(\pm 1)}] \,, \nonumber 
\end{eqnarray}
The perturbations in the $k$th eigenmode can now be written as
\begin{equation}
\begin{array}{rclrcl}
\potential &=& \hat A\, Q^{(0)} 
	\,, \quad& \curvature &=& \hat H_L\, Q^{(0)} \,, \nonumber\\
\delta\rho &=& \widehat{\delta\rho}\, Q^{(0)}\,, \quad&  \delta p   &=& \widehat{\delta p}\, Q^{(0)} \,,
\end{array}
\end{equation}
which possess only scalar components,
\begin{eqnarray}
\shift_i& =& \sum_{m=-1}^1 \hat B^{(m)}\, Q_i^{(m)}\,, \nonumber\\
v_i& =& \sum_{m=-1}^1 \hat v^{(m)}\, Q_i^{(m)}\,,
\end{eqnarray}
which possess scalar and vector components, and
\begin{eqnarray}
\shear{}_{ij} & = & \sum_{m=-2}^{2} \hat H_T^{(m)}\, Q_{ij}^{(m)}\,, \nonumber\\
\Pi_{ij} & = & \sum_{m=-2}^2 \hat \Pi^{(m)}\,  Q_{ij}^{(m)}\,, 
\end{eqnarray}
which possess all three types. Here, scalar perturbations are
denoted with superscript
$(0)$, which is elsewhere omitted. 
We will hereafter also omit the overhat in the normal mode
amplitudes since real-space objects will no longer appear. 
Thus $\hat{v}^{(0)} \equiv v$. 

\subsection{Gauge Covariant Equations}
\label{sec:eom}

The equations of motion for the matter follow from the 
Einstein equations $G_{\mu\nu}=8\pi G T_{\mu\nu}$.  
Furthermore, the Bianchi identities guarantee
$T_{\mu\nu}^{\hphantom{\mu\nu};\nu}=0$ 
which represents covariant energy and momentum conservation.

For the background, energy conservation implies
\begin{equation}
{\dot \rho}  
 = -3 {\dot a \over a}(1+w)\rho \,,
\label{eqn:rhoevol}
\end{equation} 
where $w=p/\rho$ determines the background equation of state.
Due to the isotropy of the background, momentum conservation
yields no additional constraint.  

The Einstein equation
determine the evolution of the scale factor through
\begin{equation}
\left(\dot a \over a \right)^2 
\equiv {8\pi G\over 3} a^2 \rhocr
= {8\pi G \over 3} a^2 
(\rho + \rho_S) 
\,. 
\label{eqn:hubble}
\end{equation}
Here we have divided contributions to the expansion rate into
the ordinary density and an effective density component that
does not participate in gravitational collapse and is hence 
labeled ``S'' for smooth.
The curvature provides the only component that is
smooth by fiat
\begin{eqnarray}
\rho_S & = &  -{3\over 8\pi G a^2} K\,,
\end{eqnarray}
with $w_S=-1/3$.  Even the cosmological constant is kept smooth simply
by dynamics.
However, this notation is convenient for
considering components that are approximately smooth.  
By keeping a general $\rho_S$ and $\rhoc$ here, we avoid
lengthy rederivation of the equations of motion for such
cases. Note that $\rhoc + \rho_S = \rhocr$, the so-called
critical density, and we can define a critical equation of
state $w_{cr}$ by
\begin{equation}
{\dot \rhocr} 
 =  -3 {\dot a \over a}(1+\wcr) \rhocr\,.
\end{equation}

Scalar matter perturbations obey the continuity and Euler equations
\begin{eqnarray}
\left[{d \over d\eta} + 3 {\dot a \over a}\right] \delta\rho
	+  3{\dot a \over a} \delta p
&=&
        -(\rho+p)(k v + 3\dot H_L)\,,  \label{eqn:continuity}\\
\left[ {d \over d\eta} + 4{\dot a \over a}\right] \left[(\rho + p){(v-B) \over k}\right]
&=& 
   \delta p - {2 \over 3}(1-3{K\over k^2})p \Pi \nonumber\\
&&\quad + (\rho+ p) A \,,
\label{eqn:Euler}
\end{eqnarray}
and place 2 constraints on the 4 scalar matter-variables.

The metric and matter are related by
the Einstein equations 
\begin{eqnarray}
&& (k^2 - 3K)[ H_L + {1 \over 3} H_T + {\dot a \over a}{1 \over k^2}
	( kB- {\dot H_T} )]   \nonumber \\
&& \qquad\qquad = 4\pi G a^2  \left[ \delta \rho + 3 {\dot a \over a} (\rho+p)(v-B)/k\right] \,,
\label{eqn:Poisson} \\
&& k^2 ( A + H_L + {1 \over 3} H_T ) + \left({d \over d\eta}+ 2 {\dot a \over a} \right)
	(k B - \dot H_T) \nonumber\\
&& \qquad\qquad = 8\pi G a^2 p \Pi  \,,
\label{eqn:PoissonPi}\\
&&{\dot a \over a} A - \dot H_L -{1 \over 3} \dot H_T - {K \over k^2} (kB- \dot H_T) \nonumber\\
&& \qquad\qquad=  4\pi G a^2 (\rho+p)(v-B)/k \,,
\label{eqn:Poissonzeta}\\
&&\left[2 {\ddot a \over a} - 2 \left( {\dot a \over a} \right)^2 + {\dot a \over a} {d \over d\eta}
- {k^2 \over 3}\right] A 
- \left[ {d\over d\eta} + {\dot a \over a} \right] (\dot H_L + {1 \over 3} k B) \nonumber\\
&& \qquad\qquad= 4\pi G a^2 (\delta p + {1 \over 3}\delta\rho ) \,.
\label{eqn:Poissonsynch}
\end{eqnarray}
Only two of these equations
are functionally independent.  The combination of these equations
that corresponds to the conservation equation 
$G^{\mu\nu}_{\hphantom{\mu\nu};\nu} = 0$ is automatically satisfied
by any choice of the 4 metric variables due to the
Bianchi identities.
The remaining degrees of freedom are related to 
gauge freedom as we shall see. 

Momentum conservation for vector perturbations gives the
Euler equation
\begin{eqnarray}
&& \left[{d \over d\eta}+4{\dot a \over a}\right]
[(\rho + p)(v^{(\pm 1)}-B^{(\pm 1)})/k] \nonumber\\
&& \qquad\qquad = - {1 \over 2}(1-2K/k^2)p \Pi^{(\pm 1)}\,,
\label{eqn:Eulervector}
\end{eqnarray}
and the Einstein equations give
\begin{eqnarray}
\label{eqn:Poissonvector}
&& (1-2K/k^2)(kB^{(\pm 1)} - \dot H_T^{(\pm 1)})  \nonumber\\
&&\qquad\qquad = 16\pi G a^2 
(\rho+p)(v^{(\pm 1)}-B^{(\pm 1)})/k \,,\\
&& \left[ {d \over d\eta} + 2 {\dot a \over a} \right] 
(kB^{(\pm 1)} - \dot H_T^{(\pm 1)})  \nonumber\\
&&\qquad\qquad= -8\pi G a^2 p\Pi^{(\pm 1)}\,.
\end{eqnarray}
Again the Bianchi identity reduces the number of independent equations
to two.

For the tensor modes, the Einstein equations reduce to a
single relation
\begin{equation}
\left[ {d^2 \over d\eta^2} + 2 {\dot a \over a}{d \over d\eta} + (k^2+2K) \right]
H_T^{(\pm 2)} = 8\pi G a^2 p \Pi^{(\pm 2)} \,.
\label{eqn:Poissontensor}
\end{equation}
Neither the conservation equations nor the Bianchi identity
say anything about tensor perturbations.

Although these relations are exact, 
they do not provide a closed system.  
There are in general 10 equations for 20 variables in
the background and perturbations separately.  
For the background, homogeneity and isotropy brings this to 
2 equations for the 3 variables ($a$), ($\rho,p$),
where the grouping distinguishes metric and matter categories.
For the perturbations,
the general relations are broken up into 
4 equations for the 8 variables 
($A,B,H_L,H_T$), ($\delta\rho,\delta p,(\rho+p)v,p\Pi$) for the
scalar perturbations, 2 equations for 4 variables
($B^{(\pm 1)}$, $H_T^{(\pm 1)}$), $(v^{(\pm 1)}, \Pi^{(\pm 1)})$ 
for each set of vector perturbations
and 1 equation for 2 variables
($H_T^{(\pm 2)}$), ($\Pi^{(\pm 2)}$) for each set of tensor
perturbations.
We can express the remaining $1 + 10$ degrees of
freedom as the ability to choose
the equation of state for the 
background $w = p/\rho$,
the 6 stress fluctuations 
$(\delta p, p\Pi, p\Pi^{(\pm 1)}, p\Pi^{(\pm 2)})$ 
for the perturbations, and the 
gauge (the 4 quantities ($\delta \eta$, $\delta x_i$) 
for an arbitrary coordinate shift). 

\subsection{Multicomponent Generalization}
\label{sec:multicomponent}

The conservation equations 
(\ref{eqn:continuity}), (\ref{eqn:Euler}) and 
(\ref{eqn:Eulervector}) are 
valid for each species whose stress-energy tensor is 
independently covariantly conserved.  For example, they apply
to the photon-baryon system and the dark sector which 
only interact through
gravity.
The Einstein equations (\ref{eqn:Poisson})-(\ref{eqn:Poissonsynch}),
(\ref{eqn:Poissonvector}) and (\ref{eqn:Poissontensor}) 
of course still hold with the appropriate summation over
components, e.g. $\rho= \sum_J \rho_J$.   Note that we
do not include the smooth component $\rho_S$ in the multicomponent
sum.

\section{Coordinate Choice}
\label{sec:gauge}
\subsection{Gauge Transformations}
\label{sec:gaugetransformations}

The additional four component freedom in the Einstein equations
is fixed by a choice of coordinates that relate the perturbations
to the underlying smooth background.  
The most general coordinate transformation associated with
the $k$th normal mode is \cite{Bar80}
\begin{eqnarray}
\tau &=& \tilde\tau + TQ^{(0)}\,, \nonumber\\
x_i &=& \tilde x_i + L Q_i^{(0)} + L_{\vphantom{i}}^{(1)} Q_i^{(1)} + 
	L_{\vphantom{i}}^{(-1)} Q_i^{(-1)},
\label{eqn:shift}
\end{eqnarray}
$T$ corresponds to a choice in time slicing and ($L$,$L^{(1)}$,
$L^{(-1)}$) a choice of
spatial coordinates.  Under the condition that
metric distances be invariant,
they transform the metric as \cite{KodSas84}
\begin{eqnarray}
A &=& \widetilde A - \dot T - {\dot a \over a} T\,, \nonumber\\
B &=& \widetilde B + \dot L + kT\,, \nonumber\\
H_L &=& \widetilde H_L - {k \over 3}L - {\dot a \over a} T\,, \nonumber\\
H_T &=& \widetilde H_T + kL\,, 
\label{eqn:metrictrans}
\end{eqnarray}
for the scalar perturbations and
\begin{eqnarray}
B^{(\pm 1)} &=& \widetilde B^{(\pm 1)} + \dot L^{(\pm 1)}, \nonumber\\
H_T^{(\pm 1)} &=&\widetilde H_T^{(\pm 1)} + kL^{(\pm 1)},
\end{eqnarray}
for the vector perturbations.

Similarly, they transform the components of the stress-energy tensor as\cite{KodSas84} 
\begin{eqnarray} 
{\delta\rho_J} &=& \widetilde{\delta\rho}_J - \dot\rho_J T, \nonumber\\ 
{\delta p_J} &=& \widetilde{\delta p}_J -\dot p_J T, \nonumber\\
v_J &=& \tilde v_J + \dot L, 
\label{eqn:fluidtrans}
\end{eqnarray}
for the scalar perturbations and
\begin{eqnarray}
v_J^{(\pm 1)} &=& \tilde v_J^{(\pm 1)} + \dot L^{(\pm 1)}, 
\label{eqn:fluidtransvector}
\end{eqnarray}
for the vector perturbations. All other quantities in the
metric and matter are gauge invariant.
In particular, the tensor modes do not exhibit gauge
freedom since the transverse-traceless condition on $Q^{(\pm 2)}$
is sufficient to remove the gauge ambiguity.
The gauge is thus fixed by conditions on the metric which fully
specify the transformation
($T$, $L$, $L^{(\pm 1)}$) from an arbitrary frame.

It is important to bear in mind that both the metric fluctuations ($A$, $B$, $H_L$, $H_T$) and
the matter fluctuations ($\delta\rho$, $\delta p$, $[\rho+p]v$)
take on different numerical values in different frames even 
in this covariant notation. 
For example, 
if $\rho$ evolves 
in time, a density perturbation $\delta\rho$ arises
simply from the warping of the time hypersurface on which
the perturbation is defined.  
Thus, a density perturbation differs negligibly only between
frames separated by (see Eq.~[\ref{eqn:rhoevol}])
\begin{equation}
T \ll \left[(1+w){\dot a \over a}\right]^{-1}{\delta \rho_J \over\rho_J.}
\end{equation}
The common gauge choices of the next section all agree on the
density perturbation in the clustered component
well inside the horizon.

\subsection{Gauge Choice}
\label{sec:gaugechoice}

Gauge freedom can be used to simplify the equations of motion.  
Most commonly,
it is employed to convert certain Einstein equations to
algebraic relations and/or eliminate relativistic effects from
the conservation equations.

\subsubsection{Vector Gauges}

Let us first dispose of the vector degrees of freedom.
There are two natural choices \cite{Bar80}: 
$H_T^{(\pm 1)}=0$ which fixes
the gauge completely and $B^{(\pm 1)}=0$ which leaves an arbitrary
constant offset in $H_T^{(\pm 1)}$.  The latter does not produce a dynamical
effect and can always be eliminated by specifying an initial condition
for $H_T^{(\pm 1)}$.

\subsubsection{Comoving (Scalar) Gauge}

It is useful to consider a scalar 
gauge where the metric and matter fluctuations are simply 
related \cite{Bar80}.  
Inspection of the Euler and Einstein equations shows us
that the coordinate choice $B=v$ simplifies the equations of motion 
greatly. This fixes the time slicing through $T=(\tilde v-\widetilde B)/k$.
The additional condition $H_T=0$ specifies that 
$L = \widetilde H_T/k$ and fixes the gauge completely.  
We call this the {\it comoving} gauge since here the momentum
density vanishes. 
The remaining metric variables are labeled $A = \xi$ and $H_L=\zeta$.
This choice reduces the Euler equation to the algebraic relation
for the potential
\begin{equation}
(\rho+p)\xi   = -\delta p + {2 \over 3}(1-3K/k^2)p\Pi \,,
\label{eqn:Eulercomoving}
\end{equation}
which is also simply related to the curvature through the
Einstein equation (\ref{eqn:Poissonzeta})
\begin{equation}
\dot \zetac = {\dot a \over a} \xic + 4\pi G a^2(\rho_S+p_S) \vc/k \,. 
\label{eqn:zetac}
\end{equation}
Here we have again rewritten the curvature component as a smooth
density contribution as in equation~(\ref{eqn:hubble})
for easy generalization to approximately smooth cases.

The simple relation between the metric and stress perturbation
of equations (\ref{eqn:Eulercomoving}) and (\ref{eqn:zetac})
is what makes this gauge
useful. A smooth component complicates these relations
because of the difference between a frame that is comoving with
$v$ versus the total-momentum-weighted velocity 
$v\dot\rhoc/(\dot\rhoc+\dot\rho_S)$.
With a constant comoving
curvature, the continuity equation (\ref{eqn:continuity}) reduces to 
an ordinary conservation equation
since metric changes to the fiducial volume are absent.

\subsubsection{Newtonian (Scalar) Gauge}

Finally, the {\it Newtonian} gauge is defined by $B=H_T=0$ and labels $A=\Psi$ and $H_L=\Phi$.  The gauge is completely specified
through $T=-\widetilde B/k +  \dot{\widetilde H}_T/k^2$ and
$L = - \widetilde H_T/k$.
The Einstein equations are reduced to algebraic relations
that generalize the Poisson equation of Newtonian gravity
\begin{eqnarray}
(k^2-3K)\Phi &=& 4\pi G a^2 \left[ \delta \rho + 3 {\dot a \over a}(\rho+p)v/k \right]\,,\nonumber\\  
k^2 (\Phi + \Psi) &=& -8\pi G a^2 p \Pi\,.
\end{eqnarray}
These algebraic relations and the fact that CMB anisotropies are
simply related to $\Phi$ and $\Psi$ make this gauge useful.  

\subsubsection{Gauge-Covariant Variables}

It is often useful to speak of the variables of say the comoving
gauge while in a Newtonian representation.  Bardeen \cite{Bar80}
introduced a so-called ``gauge-invariant'' language that achieves this.
We denote such techniques as gauge covariant since they amount
to introducing covariant expressions for objects that take on the
desired meaning only in a specific frame.
The only objects that cannot be made gauge covariant
are those that are ill-defined 
due to coordinate ambiguities.

To avoid confusion, we only use gauge-covariant variables
to describe metric fluctuations ($\zeta$, $\xi$, $\Phi$, $\Psi$).
Matter perturbations will always be represented in 
the comoving gauge unless otherwise specified.  Note that
$v$ is the same in comoving and Newtonian gauges.

The
comoving curvature and density can be usefully expressed
in Newtonian variables
\begin{eqnarray} 
\zetac 
        & = & \Phi + 
	2\left(\Psi-\Phi'\right) {\rhocr \over \rhoc'} \,,
\label{eqn:zetaphi}
\end{eqnarray}
\begin{eqnarray}
4\pi G a^2 \delta \rhoc &=&
(k^2-3K)\Phi\,, 
\label{eqn:hybridpoisson}
\end{eqnarray}
obtained through equations (\ref{eqn:metrictrans}) and 
(\ref{eqn:Poissonzeta}).
Likewise equation (\ref{eqn:zetac}) can be rewritten as
\begin{equation}
\zetac' - \xic = \left( \Psi - \Phi' \right) 
{\rho_S'  \over \rhoc'}\,.
\label{eqn:zetadotphi}
\end{equation}
Recall that primes represent derivatives with respect to
$\ln a$.  

As we shall see, employing both comoving and Newtonian
metric variables in the covariant language 
allows us to exploit the simple relations
to comoving stresses in the former and comoving density
perturbations in the latter. 

\section{Stress Phenomenology}
\label{sec:stressrepresentation}

\subsection{Stress Representation}
\label{sec:stresstypes}
\begin{figure}[tb]
\centerline{ \epsfxsize = 2.75truein \epsffile{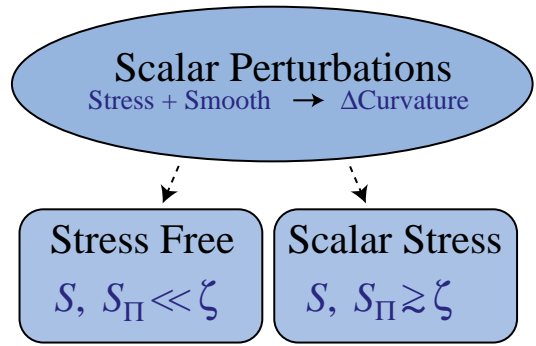}}
\vskip 0.5truecm
\caption{Scalar perturbations.  It is useful to subdivide the stress-free
class of scalar perturbations from the general possibilities.  A ``stress-free'' 
perturbation has dimensionless stresses ($\stress$,$\stressa$) that are
much smaller than the comoving curvature perturbation $\zetac$.
Note that the stress-free perturbation condition does not preclude
background or ``smooth'' stress.
}
\label{fig:scalar}
\end{figure}

\begin{figure}[tb]
\centerline{ \epsfxsize = 3.25truein \epsffile{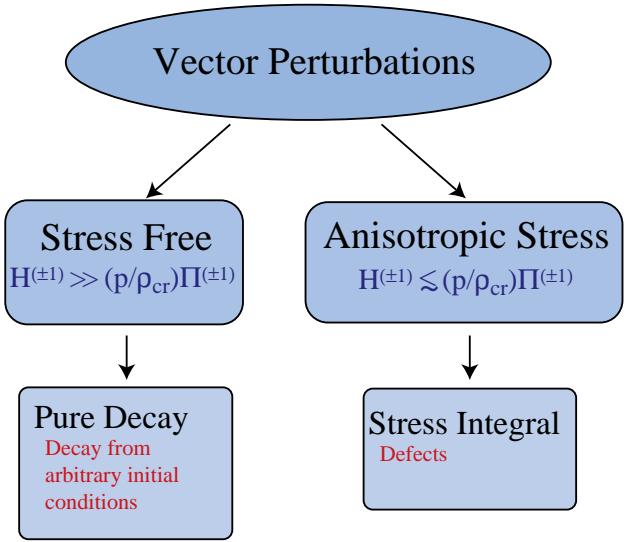}}
\vskip 0.5truecm
\caption{Vector perturbations.  Vector perturbations simply decay 
from their initial value in the
stress-free limit.  The integral solution in the presence of vector
stress is given in \S \protect{\ref{sec:vectorstress}} and applies
to defect models.
}
\label{fig:vector}
\end{figure}

\begin{figure}[tb]
\centerline{ \epsfxsize = 3.25truein \epsffile{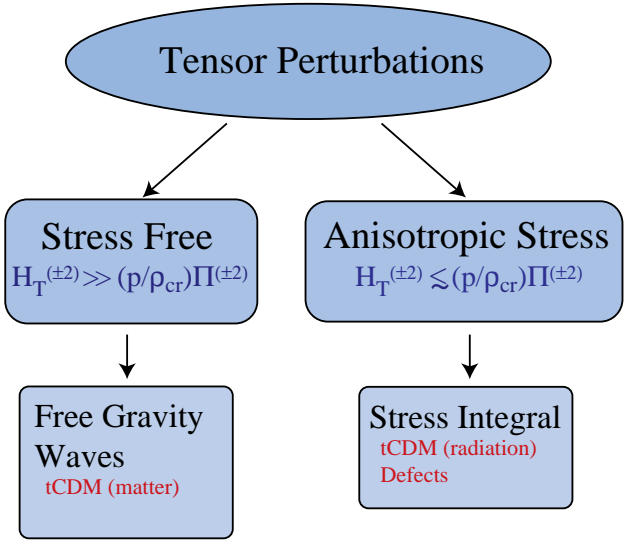}}
\vskip 0.5truecm
\caption{Tensor perturbations.  Tensor perturbations propagate as free
gravity waves in the stress-free limit as is the case of a 
matter-dominated expansion.   The integral solution in the 
presence of stresses is given in \S \protect{\ref{sec:tensorstress}}
and may be applied to propagation during radiation domination as
well as defect sources.}
\label{fig:tensor}
\end{figure}

We have seen that the stresses of the matter components completely 
determine the evolution of perturbations (see Figs.~\ref{fig:scalar},
\ref{fig:vector}, and \ref{fig:tensor}).  
The background stress is completely determined by the equation
of state $w$.  The scalar stress fluctuations are determined 
by functional relations between the pressure or isotropic
stress perturbations $\delta p$, anisotropic stress perturbation 
$p\Pi$ and the density perturbation $\delta\rho$. These 
relations may also involve hidden internal degrees of
freedom.
A model may also possess background stress without stress
perturbations and vice versa.  
We call the former a smooth stress and the latter a seed stress.
Vector and tensor 
perturbations likewise depend on the components 
$p \Pi^{(\pm 1)}$ and $p \Pi^{(\pm 2)}$
of the anisotropic stress tensor.

Scalar stress perturbations control the basic
elements of the structure formation history,
and so we pay particular attention to categorizing their properties.
As discussed in \S \ref{sec:gaugechoice}, the isotropic scalar stress poses a special
problem in that its value depends on the choice of coordinate or gauge. 
The comoving gauge (where the momentum density vanishes) provides
a  useful choice of gauge because of the simple relation 
between the metric and stress fluctuations.
We will use these coordinates
to define the {\it total} scalar stress 
\begin{equation}\label{eqn:scalarstress}
\stress = -\xi = {\delta \pc \over \rhoc+\pc} - 
	{2 \over 3}(1-3K/k^2){\pc \over \rhoc+\pc}\Pic \,.
\end{equation}

It is useful to isolate gauge invariant aspects of the stress
perturbation.
The {\it anisotropic} scalar stress
\begin{equation}
\stressa = -8\pi G a^2 \pc \Pic/k^2 \,,
\end{equation}
is gauge invariant by definition; 
this form of the anisotropic stress also enters 
into the Einstein equations separately from $\stress$.

An adiabatic stress perturbation obeys
\begin{equation}
\delta p_A = (\rhoc+\pc)\stressad  = {\pc' \over \rhoc'}
\delta \rhoc\,.
\label{eqn:adiabaticstress}
\end{equation}
Although $\delta p_A$ is not gauge invariant, the adiabatic
sound speed is 
\begin{equation}
c_s^2 \equiv
{\delta p_A \over \delta \rhoc} = {\pc' \over \rhoc'} \,. 
\label{eqn:sonicstress}
\end{equation} 
This gauge invariance implies that there are
no coordinate ambiguities when discussing the pressure support
of adiabatic fluctuations.  

The remaining gauge-invariant pressure perturbation is 
\begin{equation}
\stressna = {1 \over \rhoc + \pc} \left( \delta \pc -
	{\pc' \over \rhoc'} \delta \rhoc \right)\,.
\label{eqn:nonadiabaticstress}
\end{equation}
Unlike adiabatic stresses,
these may be present even when the comoving 
density perturbation $\delta \rhoc$
is negligible (see Eq.~[\ref{eqn:hybridpoisson}]).  Entropic
stresses are the primary means of structure formation in most
isocurvature models.

The comoving-gauge analogue of $\stressad$ and $\stressna$
are also useful.
As long as $\delta \pc/ \delta \rhoc$
is less than unity, superhorizon stresses
are negligible compared with curvature fluctuations.  
We call the part of $\delta \pc/\delta\rhoc$
that is separable
in time and space the {\it comoving} sound speed
$c_C^2 = f(k) g(\eta)$ and
the accompanying stress {\it sonic} 
\begin{equation}
\stresssonic = c_C^2 {\delta \rhoc \over \rhoc + \pc} \,.
\end{equation}
Separability
is not a 
gauge-invariant property, but
this is not in itself a problem because the comoving frame
is dynamically special.   

We call the remaining isotropic stress
the entropic stress
\begin{equation}
\stressg = {1 \over \rhoc + \pc} \left( \delta \pc -
	c_C^2 \delta \rhoc \right)\,,
\label{eqn:entropicstress}
\end{equation}
such that the total stress is 
\begin{equation}
\stress = \stresssonic+\stressg+
{2\over3}{k^2-3K\over 8\pi Ga^2(\rhoc+\pc)}\stressa \,.
\end{equation}
Note that if the comoving sound speed equals the adiabatic
sound speed $c_C^2=c_s^2=p'/\rho'$ then $\stresssonic =
\stressad$ and $\stressg = \stressna$.

\subsection{Seed Stress}

Seed stresses provide a special case with unique properties.
The effect of
seed perturbations are in fact simpler to understand than 
those of the fluid type because the problem decouples 
completely.  If the seeds do not interact directly with
other types of matter, the conservation equations
(\ref{eqn:continuity}) and (\ref{eqn:Euler}) imply that
the metric perturbations only affect the seed 
perturbations at second order since bare $\rhoseed$ and $\pseed$
terms
may be dropped.  They become \cite{VerSte90}
\begin{eqnarray}
\left[{d \over d\eta} + 3 {\dot a \over a}\right] \delta\rhoseed
&=&
        -k(\rhoseed+\pseed)\vseed
-  3{\dot a \over a} \delta \pseed\,,
\label{eqn:continuityseed}\\
\left[ {d \over d\eta} + 4{\dot a \over a}\right] (\rhoseed + \pseed)
	{\vseed \over k}
&=& 
   \delta \pseed - {2 \over 3}(1-3K/k^2)\pseed \Piseed \,.
\label{eqn:Eulerseed}
\end{eqnarray}
The basic principles of how stress fluctuations affect the
gravitational potential still hold but here there is a stress
contribution that is truly external to the system of metric
fluctuations.    
It is again possible to have large-scale 
entropic stress perturbation
in the absence of initial curvature or density perturbation.  

The formal solution to equations~(\ref{eqn:continuityseed}) and
(\ref{eqn:Eulerseed}) are
\begin{eqnarray}
(\rhoseed + \pseed) \vseed /k & = & a^{-4} \int d\eta a^4 [\delta \pseed
-{2 \over 3}(1-3K/k^2)\pseed\Piseed]  \,,\nonumber\\
\delta\rhoseed &=& -a^{-3} \int d\eta a^3[ k(\rhoseed+\pseed)\vseed
	+ 3 {\dot a \over a} \delta\pseed ]\,.
\end{eqnarray}
The task
of understanding a seed model like defects reduces to understanding
its stresses, but this is a formidable task in realistic seed models
such as cosmological defects (e.g.~\cite{PenSelTur97}). 

\subsection{From Stresses to Curvature}

These stresses are the fundamental sources and sinks of the
metric perturbations.  We therefore seek to express 
the comoving and Newtonian curvatures in terms of the stresses.
Combining equations (\ref{eqn:zetaphi}),
(\ref{eqn:zetadotphi}), and (\ref{eqn:scalarstress}) with
\begin{equation}
\Psi = -\Phi + \stressa \,,
\end{equation}
yields
\begin{equation}
\zetac' = -\stress + [\Phi' + \Phi -\stressa]
\left({{\rhocr'\over \rhoc'} -1}\right)\,,
\label{eqn:zetageneral}
\end{equation}
and
\begin{equation}
{\sqrt{\rhoc} \over a} \left[ {a \over \sqrt{\rhoc}} \Phi \right]'
        = 
	-{1\over2}{\rhoc'\over\rhoc}\zetac + \stressa\,.
\label{eqn:phideriv}
\end{equation}
In a non-critical (non-flat, $\rhoc\ne\rhocr$) universe, we combine these
to eliminate $\zetac$ from the evolution equation
for the Newtonian curvature, giving
\begin{eqnarray}
&&\Phi'' + \left( 1 - {\rhoc'' \over \rhoc'} + {1 \over 2} {\rhocr' 
	\over \rhocr} \right) \Phi' 
	+ \left( {1 \over 2} {{\rhocr' +\rhoc'}\over \rhocr} 
	- {\rhoc'' \over \rhoc'} \right)
	\Phi  \nonumber\\
&&\qquad=
	{1 \over 2} {\rhoc' \over \rhocr} \stress
	+ \stressa' + \left({1 \over 2} {\rhocr' + \rhoc' \over \rhocr} - 
	{\rhoc'' \over \rhoc'}\right) \stressa \,.
\label{eqn:phigeneral}
\end{eqnarray}
Recall that the Newtonian curvature is
simply related to the comoving density perturbation through equation
(\ref{eqn:hybridpoisson}).

In a critical-density universe ($\rhoc=\rhocr$), 
equations (\ref{eqn:zetageneral}) and (\ref{eqn:phideriv}) can be formally
solved as integrals over the stress fluctuations to yield
\begin{equation}
\zetac(a) = \zetac(0) -\int {da \over a} \stress \,,
\label{eqn:zetaformal}
\end{equation}
and 
\begin{equation}
\Phi= \zetac - {\sqrt{\rhoc} \over a} \int { da \over \sqrt{\rhoc}} 
\left[ \zetac - \stress - \stressa \right] + C {\sqrt{\rhoc} \over a}\,.
\label{eqn:phiformal}
\end{equation}
The last term is the decaying mode of $\Phi$ where $C=$const.

There are three general conclusions that we can draw from
equations~(\ref{eqn:zetaformal}) and (\ref{eqn:phiformal}).
The first is that in the absence of an initial comoving 
curvature perturbation ($\zetac(0)=0$), a stress fluctuation will
generate one of order $\zetac \sim -S$.  The same goes for the
Newtonian curvature $\Phi \sim -S$.  The reason for this behavior
is that a stress gradient $k\delta \pc$ generates a potential flow
with $(\rhoc+\pc)\vc \sim (k\eta)\delta \pc$ which generates
a density perturbation of $\delta \rhoc \sim -(k\eta)^2 \delta \pc$
and hence a curvature perturbation of $\Phi \sim -\delta \pc/\rhocr$.
Note this intuitive argument fails for other gauge choices. 

Second, starting with a curvature perturbation and assuming
sonic stresses $\delta p = c_C^2 \delta \rho$, 
it is clear that the same mechanism
of generating flows will generate an opposing curvature fluctuation
$\Delta \Phi \sim -c_C^2 \delta\rho/\rho \sim -(c_C k\eta)^2 \Phi$
that will destroy the initial curvature fluctuation 
when $c_C k\eta \sim 1$.  
In physical terms this occurs because pressure support prevents
perturbations from collapsing and hence causes the curvature
perturbation to redshift away.

Finally, the anisotropic stress contributes
to the total stress $S$ and thus can both create
and destroy comoving curvature fluctuations.  
Furthermore, it enters 
separately into the Newtonian curvature through
equation~(\ref{eqn:phiformal}).  
This is because the Newtonian frame, unlike the comoving frame,
is defined to be globally shear free ($B=H_T=0$). 
The coordinate transformation
that maps the comoving frame to the shear free frame depends on
the anisotropic stress and hence aliases background evolution into
contributions to the Newtonian curvature.

Unfortunately, despite their general appearance, equations 
(\ref{eqn:zetaformal}) and (\ref{eqn:phiformal}) 
are only formal solutions
since the time evolution of the stress sources generally
depend on the 
curvature fluctuation itself.  We will use the special 
properties of smooth, anisotropic, entropic, and sonic
stresses to address this problem in
\S \ref{sec:stressfree}, 
\S\ref{sec:purestress}, and \S\ref{sec:mixedstress}.

\subsection{From Curvature to Observables}
\label{sec:curvtoobs}

As discussed in \S \ref{sec:generalphenomenology}, the Newtonian
curvature is directly related to observables in the CMB and large
scale structure.  
We are now in a position to quantify these relations.
The Newtonian curvature $\Phi$ and potential
$\Psi$ encapsulate all observable properties of scalar fluctuations.
The contribution of a given $k$-mode to the amplitude of 
$\ell$th multipole moment of the CMB anisotropy 
is given by \cite{HuSug95}
\begin{eqnarray}
{\Theta_\ell \over 2\ell+1} &\approx& \int_0^{\eta_0} d\eta
e^{-\tau} \Big\{ \left[ \dot\Psi -\dot\Phi + 
	\dot\tau (\Theta_0+\Psi) \right] \nonumber\\
&& \times
	j_\ell[k(\eta_0-\eta)] + \dot\tau v_b j_\ell'[k(\eta_0-\eta)]
	\Big\} \,,
\label{eqn:thetaell}
\end{eqnarray}
where we have dropped the small correction due to the polarization
of the CMB. $\tau$ is the optical depth to Compton scattering
between the present ($\eta_0$) and the epoch in question ($\eta$);
$\Theta_0$ is the photon temperature perturbation in Newtonian gauge,
and $v_b$ is the baryon velocity in Newtonian or comoving gauge.  
For an open universe, the spherical Bessel function
$j_\ell$ is replaced by the hyperspherical bessel function.  Note
that the prime here and here only 
refers to a derivative with respect to the
argument of the Bessel function.
With the random phase assumption for the $k$-modes, the scalar 
contribution to power spectrum of
the anisotropies is 
\begin{equation}
C_\ell = {2 \over \pi} \int {dk \over k} k^3 {\left<\Theta_\ell^* 
	\Theta_\ell\right> \over ({2\ell+1})^2} \,. 
\label{eqn:cl}
\end{equation}

In equation (\ref{eqn:thetaell}), the $\dot \Psi -\dot\Phi$ 
term leads to the so-called integrated Sachs-Wolfe (ISW) effect
and contributes once the optical depth to scattering becomes small,
i.e.~after last scattering.  The other terms, $\Theta_0+\Psi$
and $v_b$, are localized to the last scattering surface itself
and represent the effective temperature of the distribution
and the Doppler effect respectively.  
They are responsible for the
so-called acoustic peaks in the CMB spectrum, the
morphology of which can be directly read off of the 
metric driving terms \cite{HuSug95}.

The influence of time variability in $\Psi$ and $\Phi$
depends on how the variation rate $1/\Delta\eta$ compares
with the perturbation crossing time for
sound (before
last scattering) and light (after last scattering).
For variations on a much shorter time scale  ($k\Delta\eta\ll 1$ or 
$kc_s\Delta\eta \ll 1$), only the total change 
$\Delta(\Psi-\Phi)$ is observable
since all photons suffer a uniform gravitational redshift.
For variations on a much longer timescale the effects 
mainly cancel out as the photons either traverse many wavelengths
of the fluctuation during the variation with redshifts and
blueshifts cancelling or equivalently undergo many acoustic
oscillations. 
Changes whose duration 
are synchronized
with the oscillation period are the most effective.  
These general considerations apply to metric variations of 
the vector and tensor type as well.
Two commonly encountered examples are the cancellation cut off
at high $\ell$ for a uniform potential decay and the driving
of acoustic oscillations from synchronized potential decay in
the radiation dominated epoch.

A constant potential also leads to observable effects through
the effective temperature term $\Theta_0+\Psi$.
Since $\Theta_0$ is the temperature perturbation in Newtonian 
gauge, it can be obtained by a
gauge transformation from the comoving gauge  
by noting that in that gauge both the density perturbation 
and the potential $\xi$ are negligible because stress perturbations
must be negligible for the potential to remain constant.
Equation~(\ref{eqn:metrictrans}) and (\ref{eqn:fluidtrans}) then imply
\begin{eqnarray}
\Theta_0 &=& {\delta \rho_\gamma \over 4\rho_\gamma}
	-{\dot a \over a} v/k \,, \nonumber\\
\Psi &=& \dot v/k + {\dot a \over a} v/k \,.
\end{eqnarray}
For adiabatic fluctuations, $\delta \rho_\gamma /\rho_\gamma 
\propto \delta\rhoc/\rhoc$ and is hence negligible outside the
horizon by virtue of the Poisson equation (\ref{eqn:hybridpoisson}). 
Since $\Psi$ is a constant by assumption, 
these equations can be integrated to give
\begin{equation}
\Theta_0  =  - {2 \over 3(1+w)} \Psi \,,
\label{eqn:effectivetemp}
\end{equation}
and hence $\Theta_0+\Psi = \Psi (1+3w)/(3w+3)$. 
This relation ultimately comes from the fact that $\Psi$
represents a time shift and $a \propto t^{2/3(1+w)}$ \cite{WhiHu97}.
For $w=0$, this
reduces to the well-known result that the effective temperature is 
$\Psi/3$ in the adiabatic sCDM model \cite{SacWol67}.  
Compared with this model, those that are dominated by the ISW
term $\Delta(\Psi-\Phi) \approx 2\Psi 
- \stressa$ like traditional isocurvature 
and smooth models potentially have up to 6 times the anisotropy
for a given potential fluctuation.

The behavior of the density perturbations that underlies 
large-scale
structure are even more directly related to the Newtonian 
curvature perturbation.  Inside the horizon, all reasonable
choices of gauge agree on the density perturbation.  In
particular, the comoving gauge density perturbation 
is algebraically related to $\Phi$  by 
equation~(\ref{eqn:hybridpoisson}) and hence allows a
simple translation of results for one to the other.
Consequently, the relation between temperature and potential
fluctuations discussed in the last paragraph
translates directly into a relation between density perturbations
and CMB anisotropies.

With this relation, the
so-called transfer function of the density
perturbations
below the current horizon can be simply read off
of the time history in the potential without the usual gauge
ambiguity in defining the initial density perturbation,
\begin{equation}
T(k) = {\Phi(\eta_0,k) \over \Phi(0,k)}
       {\Phi(0,0) \over \Phi(\eta_0,0)}\,.
\end{equation}  
The power spectrum of density perturbations today is then
proportional to $T(k)^2 P_{\rm initial}$.  Any process that makes
the potential decay relative to the $k=0$ mode will produce a
downturn in $T(k)$ and a reduction of small-scale power. 

\begin{figure}[t]
\centerline{ \epsfxsize = 2.5truein \epsffile{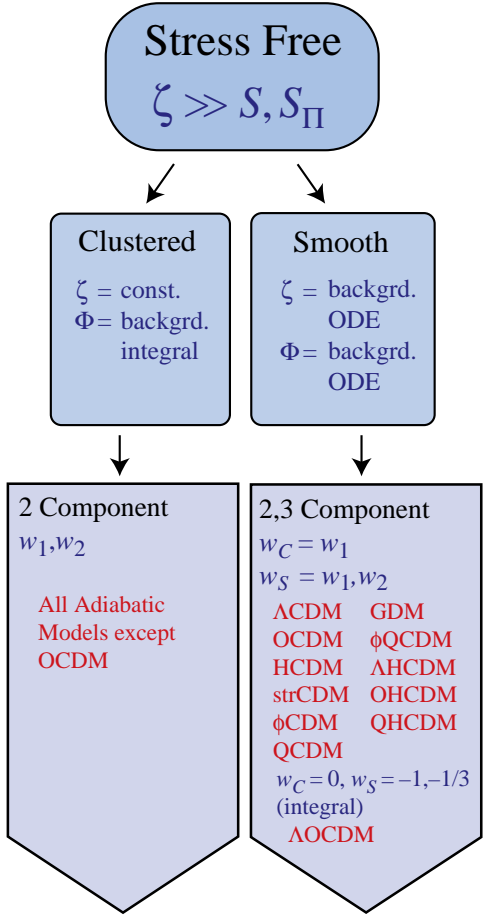}}
\vskip 0.5truecm
\caption{Stress-free scalar perturbations.  If all of the matter
which drives the expansion participates in gravitational instability,
the matter is said to be ``clustered'' otherwise there exists
a ``smooth'' component.  In either case, the perturbations depend
only on the background (``backgnd.'') equation of state $w$ and can
be tracked with simple techniques (middle row), which lead to
exact solutions (bottom row) that describe behavior in 
a wide range of models. 
} 
\label{fig:stressfree}
\end{figure}

\section{Stress-Free Perturbations}
\label{sec:stressfree}

Adiabatic models for structure formation, those whose initial conditions contain
true comoving curvature perturbations, 
all go through a period in which the scalar stress perturbation may be
neglected.
This is a direct consequence of causality.  Stress
gradients affect structure formation simply by the causal motion
of matter.  On scales larger than the horizon, 
the change in 
the comoving curvature perturbations due to 
causal motion always can always be neglected.  

We begin 
in \S \ref{sec:stressfreeclustered} 
with adiabatic models that have no smooth components.  
Although the comoving curvature
remains at the value set by the initial condition, 
the Newtonian curvature
depends on the equation of state.
We consider then
in \S \ref{sec:stressfreegeneral} the 
effect of a smooth component on the
Newtonian curvature (and hence the density perturbation).
For
each case, we begin with a general description of the resultant
phenomenology.  We then illustrate the phenomenology with
full solutions of the perturbation equations and discuss 
applications within the current generation of
structure formation models of \S \ref{sec:currentzoo}.  We
will follow this pattern throughout the paper.  An overview
of results is given in Fig.~\ref{fig:stressfree}.

\subsection{Clustered Case}
\label{sec:stressfreeclustered}

We begin with the case in which
there is no 
smooth component ($\rho_S=0$) and stress fluctuations are negligible
compared with metric fluctuations ($\stress \ll \zetac$).
Here,
equation~(\ref{eqn:zetageneral}) simply implies
\begin{equation}
\zetac = {\rm const.}
\end{equation}
Equation (\ref{eqn:phiformal}) then gives
\begin{equation}
\Phi = \zetac \left(1 - {\sqrt{\rhoc}\over a}\int{d a\over\sqrt{\rhoc}}\right)
	+ {\sqrt{\rhoc}\over a}\int{d a\over\sqrt{\rhoc}} \stressa
	+ C{\sqrt{\rhoc} \over a}.
\label{eqn:phicluster}
\end{equation}


We will focus only on the behavior of the first term.
The second term may be neglected if $\stressa\ll\zetac$.
However, unlike $\stress \ll \zetac$, $\stressa \ll \zetac$ is not 
a consequence of causality and is mildly violated in the case
of free radiation \cite{HuSug95}
and can be strongly violated in defect
models \cite{DurKun98}.  We return to consider its effects in
\S \ref{sec:anisotropic}.
The last term is a decaying mode that carries no comoving
curvature ($\zetac=0$).  For a constant equation of
state $\wc$, it scales as
\begin{equation}
\Phi \propto a^{-3(1+\wc)/2 - 1}.
\end{equation}

It is apparent from the form of the first term that since $\rhoc$
cannot grow with time ($\wc \ge -1$), 
the first integral in equation (\ref{eqn:phicluster})
goes to a constant between $0$ and $1$ with the
two extremes representing $\wc \rightarrow \infty$ and 
$\wc=-1$ respectively.  
The integral is dominated by the most recent
epoch; only the equation of state at the epoch in question
matters.  
It is simple to show that during periods where 
$\wc$ is approximately constant \cite{Lyt85},
\begin{equation}
{\Phi \over \zetac} \rightarrow { 3 + 3 \wc \over 5 + 3 \wc}\,.
\label{eqn:phizetac} \end{equation} 
At each epoch where $\wc$ decreases from $w_1$ to $w_2$, 
the Newtonian curvature decreases by 
\begin{equation}
{\Phi(w_1)-\Phi(w_2) \over \Phi(w_1)} 
= 2 {(w_1 - w_2)  \over (1+  w_1)(5+3 w_2)} \,. 
\end{equation}

These results are easy to understand in the non-relativistic
limit.
In the absence of stresses, the gradients in
the gravitational potential set up flows as $v \sim (k\eta) \Phi$.  The divergence
of this flow generates density perturbations $\delta \sim (1+\wc)
(k\eta)^2 \Phi$ for constant $w$.  By the
Poisson equation $\Phi \sim (k\eta)^{-2} \delta$, this is exactly the rate of
growth needed to keep the potential constant.  Here we have used the fact that
$4\pi G a^2 \rhoc = 3 (\dot a / a)^2 /2 \sim \eta^{-2}$ 
if there are no smooth components.
When $\wc$ decreases from $w_1$ to $w_2$,  the change affects
the proportionality constants but not the scaling between
$\delta$ and $\Phi$.  This leads to a decrease in the Newtonian 
potential to a new constant.

{\it Full Solutions.\ ---}
If the matter is predominantly composed of two
components with different but constant equations of
state $w_1$ and $w_2$,
we can solve equation (\ref{eqn:phicluster}) exactly provided
$\stressa \ll \zetac$. 
The result is
\begin{eqnarray}
\label{eqn:phiclusteredsoln}
{\Phi\over \zetac} &=& 1-{2\over5+3w_1}
	F\left[1,{1\over2};1+n;{y \over 1+y}\right]\\
     &=& 1-{1 \over 3}{1 \over w_1 - w_2} (1+y)^{1/2} y^{-n}
        B_{y/(1+y)}\left(n,{1\over2}-n\right) \,,\nonumber
\end{eqnarray}
where $n \equiv (5+3w_1)/(6w_1-6w_2)$ and
\begin{equation}
y=\rho_2/\rho_1 \propto a^{3(w_1-w_2)}.
\end{equation}
Here, $B_x(p,q)$ is the incomplete beta function and $F(a,b;c;x)$ is
Gauss's hypergeometric function (i.e.\ ${}_2F_1$).
The combination $y/(1+y)$ often enters into such solutions as it is
just $\rho_2/\rhocr$, the fractional density perturbation supplied
by component 2 as a function of time.

Any case with $2n$ equal to an integer can be expressed
in elementary form.  In particular, if $n = N$ or $N+1/2$ for an
integer $N\ge0$, then
\begin{eqnarray}
&& F\left(1,{1\over2}; n+1; x\right) \\
&&\quad=
{1\over x}\sum_{k=0}^{N-1}{\Gamma(n+1)\Gamma(n-k-1/2)\over
        \Gamma(n-k)\Gamma(n+1/2)}\left(x-1\over x\right)^k\nonumber\\
&&\qquad  + \left(x-1\over x\right)^{N}{\Gamma(n+1)\Gamma(n-N+1/2)
\over
        \Gamma(n-N+1)\Gamma(n+1/2)}f_n, \nonumber
\end{eqnarray}
where
\begin{equation}
f_n = \cases{1/\sqrt{1-x},&$n=N$;\cr
        {1\over2\sqrt{x}}\log{1+\sqrt{x}\over1-\sqrt{x}}, & $n=N+{1\over2}$.\cr}
\end{equation}

\begin{figure}[tb]
\centerline{\epsfxsize=3.6truein \epsffile{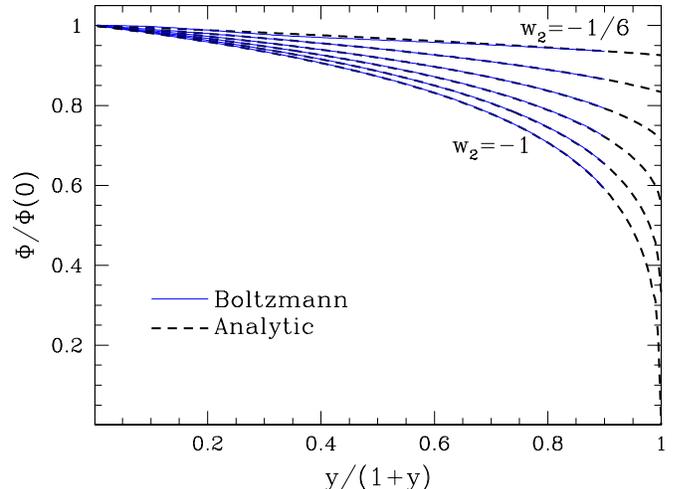}} 
\caption{Stress-free clustered case.  In the case where all contributions to the
expansion rate also cluster, the comoving curvature is constant but the
Newtonian curvature changes with the equation of state $w$.  For the case of the 
transition between $w_1=0$ and $w_2$ between $-1$ and $-1/6$, we show 
here the analytic solution of equation \protect{(\ref{eqn:phiclusteredsoln})} is
compared with a full numerical Boltzmann solution of a QCDM model (including radiation)
for a mode that is outside the horizon at the given epoch.   
Notice that only for $w_2=-1$ (cosmological
constant) does the potential decay to zero as $\rho_2/\rhocr = y/(1+y) \rightarrow 1$.}
\label{fig:phic}
\end{figure}

{\it Applications.\ ---}
These solutions apply to any adiabatic model with a matter-dominated epoch and may
be used to explore the behavior of perturbations entering and exiting the
matter-dominated epoch.    For example, the matter-radiation case reduces to \cite{KodSas86}
\begin{eqnarray}
{\Phi \over \zetac} &=& 
{3\over5} + {2\over15y} - {8\over15y^2} - {16\over15y^3}
+ {16\sqrt{1+y}\over15y^3}\,.
\end{eqnarray}
This solution only  holds in the absence of anisotropic stress perturbations
and does not strictly apply to the usual radiation components.
The neutrinos carry anisotropic stress \cite{HuSug95} as do the photons after
recombination.  We consider how such effects can be taken into account
in \S \ref{sec:anisotropic}.

For adiabatic models where there exists a component with 
$w_2 < 0$ \cite{QChicago,QPaul,Hu98}, these solutions describe 
the exit from the 
matter-dominated epoch.  A comparison of the analytic solution with
full numerical solutions in those cosmologies is given in 
Fig.~\ref{fig:phic}.
A special case is $w_1=0$ and $w_2=-1$, the matter to cosmological constant 
transition, where \cite{Bil92,Mat95}
\begin{eqnarray}
{\Phi\over \zetac} 
	= 
	{3 \over 5}
	\left[ 1 - {1 \over 3}(1+y)^{1/2}y^{-5/6} B_{y/(1+y)}(5/6,-1/3)\right]\,.
\end{eqnarray}
As is evident from equation~(\ref{eqn:phizetac}),
the cosmological constant is the only case where
the 
Newtonian
curvature decays to zero.
Note in particular that $w_2=-1/3$ does not correspond to the
behavior of spatial curvature
in spite of the fact that the effect on the expansion rate is the same.   The presence
of fluctuations in the $w_2$ component prevent the gravitational potential from 
decaying to zero.  Across the $w_1=0$ (matter-dominated) to 
$w_{2}=-1/3$ transition,
the Newtonian curvature goes from $3\zetac/5$ to $\zetac/2$.
Thus, the string-dominated (strCDM) model
which has $w_{\rm str}=-1/3$
does not behave like an open model on the large scales relevant
for the CMB (c.f. \cite{SpePen97}) 

\subsection{Smooth Components}
\label{sec:stressfreegeneral}

We next consider how smooth components affect the
growth of structure. 
Although curvature is the only component that is smooth
by definition and a cosmological constant the only one that is smooth
by dynamics, under certain 
circumstances other components can be approximately smooth.

We define a component ($S$) to be smooth if its density fluctuations
are small in comparison to those of the clustering components ($C$),
i.e.
$\delta\rho_S \ll \delta \rhoC$
regardless of whether $\rho_S < \rhoC$.  
To be maintained dynamically,
the respective energy fluxes must also satisfy 
$(\rho_S + p_S)v_S \ll (\rhoC + \pC)\vC$. 
In this section,
we add the subscript ($C$) on the remaining matter 
to remind the reader that part of the total matter density 
has been designated as effectively smooth in the division 
of equation~(\ref{eqn:hubble}).   
Our analysis also applies to cases 
where $\bar{\delta\rho_S} \ll \delta \rho_C$
where the time-average is over the dynamical time of
the $C$ component (see \ref{sec:entropygen}).

One cannot demand that $\delta\rho_S$
be identically zero since the continuity equation
(\ref{eqn:continuity}) generates a density fluctuation as
the metric curvature changes unless $w_S \equiv p_S/\rho_S 
=-1$ (a cosmological constant).                       
The fractional density fluctuation is generically at least
of order the curvature fluctuation $\Phi$.  Since this term
exceeds the energy flux term outside the horizon 
due to causality, energy conservation forbids smooth components
(with $w_S\ne 1$) on these scales. 

Components that are smooth within the horizon are possible.
For example,
the $C$ component may be 
driven to collapse by potential gradients
while the $S$ component is supported against collapse by 
stress gradients (see in \S \ref{sec:entropygen}).
Here $\delta \rho_S/\rho_S \sim \Phi \ll \delta \rhoC/\rhoC$.
Since these stress gradients are set up exactly so as to keep 
the component smooth, one can replace such stress effects in
equation~(\ref{eqn:phigeneral}) with an additional 
smooth density component \cite{TurWhi97}.
The remaining perturbations can then be approximated as 
stress-free and generate curvature fluctuations as
\begin{equation}
\Phi'' + \left( 1 - {\rhoC'' \over \rhoC'} + {1 \over 2} {\rhocr' 
	\over \rhocr} \right) \Phi' 
	+ \left( {1 \over 2} {{\rhocr' +\rhoC'}\over \rhocr} 
	- {\rhoC'' \over \rhoC'} \right)
	\Phi = 0 \,.
\label{eqn:phismoothclustered}
\end{equation}
Now even for constant $\wC$, $\Phi'=0$ no longer solves the equation
of motion.  Mathematically, $\rho_S$ adds to the
expansion drag ($\Phi'$) terms but not the gravitational
($\Phi$) terms.  Physically, 
potential flows still create density perturbations as
$\rhoC \sim (\rhoC+\pC) (k\eta)^2 \Phi$ but the
Poisson equation leads to a smaller potential
$\Phi = (k\eta)^{-2} \delta\rhoC/(\rhoC+\rho_S)$. As this 
process continues, $\Phi$ decays away.  

Equation~(\ref{eqn:phismoothclustered}) 
has simple solutions in the limit
that $\rhocr' \gg \rhoC'$, as is usually the case when the smooth component dominates
the expansion.  In this case, the general solution to the equation is
\begin{equation}
\Phi = C_1 a^{-1} + C_2 a^{-1} \int d\ln a {a \rhoC' \over \rhocr^{1/2}} \,.
\label{eqn:phismoothregime}
\end{equation}
Both terms represent decaying modes as long as $\wcr - 2 \wC < 1$.  

{\it Full Solutions.\ ---}
The full solution to equation~(\ref{eqn:phismoothclustered}) can be
obtained analytically for a clustering
component with a constant equation of state and present-day 
fractional density contribution pair ($w_1$,$\Omega_{C1}$), 
a smooth component 
with ($w_1$,$\Omega_{S1}$), and/or
a second smooth component with ($w_2$,$\Omega_{S2}$):
\begin{eqnarray}
\label{eqn:ultimateequation}
\Phi_{j} 
&=& y^{1/3\Delta w} \left( { y \over 1+ y} \right)^{\alpha_j} \\
&& \times        F\left(\alpha_j, \alpha_j+{1\over2}; 
	2\alpha_j+1-{3w_1+1\over6\Delta w}; {y \over 1+y}\right),\nonumber
\end{eqnarray}
where $\Delta w = w_2-w_1$ with
\begin{eqnarray}
\alpha_j &\equiv& {1+3w_1\over12\Delta w}
	\left[1\mp\sqrt{1+24{\Omega_{C1}\over {\Omega_{C1}+\Omega_{S1}}}
	{1+w_1\over(1+3w_1)^2}}\right],\nonumber\\
y &=& \rho_2 / \rho_1 = {\Omega_{S2} \over 
	\Omega_{S1}+\Omega_{C1}} a^{-3\Delta w},
\end{eqnarray}
where $j=1,2$ for the growing and decaying modes corresponding to
$-,+$ in the $\alpha_j$ equation.
Note that negative contributions to the critical density, e.g.
positive spatial curvature $K>0$, are also covered by these solutions.

The hypergeometric solution allows one to identify elementary solutions
more easily.  In particular, cases where $\alpha_j=-|N_1/2|$, 
$(3w_1+1)/6\Delta w = N_1$, or $2\alpha_j - (3 w_1+1)/6\Delta w + 1/2
= N_1$  
can be expressed in terms of elementary functions.   Cases
of the form 
$F(a,b;b+N_1;x)$, 
$F(a,b;a+N_1;x)$, 
or $F(N_1/2, N_2/2; N_3/2;x)$
can be expressed in terms of elementary functions, incomplete
beta functions, and/or complete elliptic integrals.
Here, the $N_j$ are integers and $a$ and $b$ are real numbers.
For example, the case of $w_1=0$ and $w_2=-1/6$ can be expressed in 
closed form for any $\Omega_{C1}/(\Omega_{C1}+\Omega_{S1})$:
\begin{equation}
\Phi_j \propto {y^{\alpha_j}\over a}
        {2\alpha_j+1+\sqrt{1+y}\over (1+\sqrt{1+y})^{2\alpha_j+1}}\,.
\end{equation}
For $w_{1}=0$, any case in which $2\alpha_j+1/6w_{2}+1/2$ is an integer
can be simplified.  One can also 
simplify the growing mode of cases with
$w_1 = \Omega_{S1}=0$ and $w_2^{-1}$ equal to an integer.

\begin{figure}[tb]
\centerline{\epsfxsize=3.6truein \epsffile{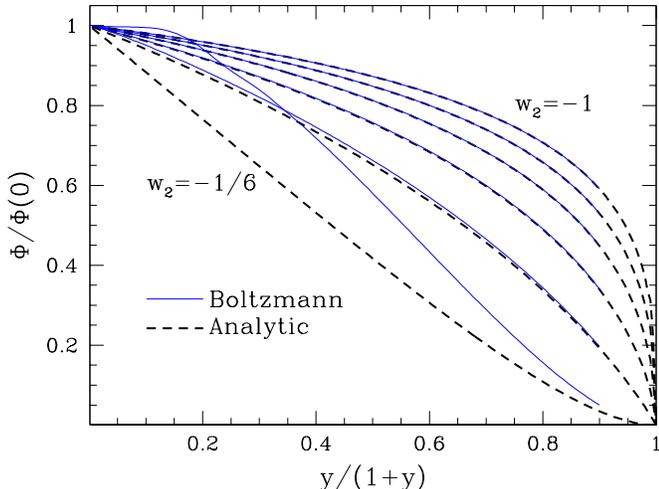}} 
\caption{Single smooth component.  The same as Fig.~\protect{\ref{fig:phic}} except
that $w_2$ is taken to be a smooth component.  Here equation (\protect{\ref{eqn:ultimateequation}})
is compared to a QCDM model but for a mode that is well inside the horizon at $y=1$.
Discrepancies at early times are due to radiation contributions in the QCDM models
and are particularly pronounced for $w_2 \rightarrow 0$ since then the epoch when the
Q and matter components are equal is driven into the radiation dominated era for this flat $\Omega_m=0.35$ model. }
\label{fig:phis}
\end{figure}

Finally for completeness, there is a well-known special case of
equation~(\ref{eqn:phismoothclustered}) that is not completely
covered by
equation~(\ref{eqn:ultimateequation}) but that does have an integral solution.  
This involves a clustering
component $w=0$ (CDM) and a smooth component composed of an arbitrary
admixture of $w=-1/3$ (curvature) and $w=-1$ ($\Lambda$) pieces \cite{Hea77}: 
\begin{equation}
\Phi \propto  a^{-1}\rhocr^{1/2} \int {d\ln a \over a^2 \rhocr^{3/2}}, \qquad
\Phi \propto a^{-1}\rhocr^{1/2} \,,
\label{eqn:lambdacurvature}
\end{equation}
for the growing and decaying modes respectively.  Of course,
if either the curvature or $\Lambda$ component is negligible,
the solutions have an analytic form described by 
equation (\ref{eqn:ultimateequation}).  Unfortunately, this solution 
cannot be generalized to arbitrary combinations of smooth components since
it relies on the fact that $w=-1/3$ does not 
accelerate the expansion and that $w=-1$ gives a 
constant density contribution.

{\it Applications.\ ---} Smooth components 
are widely found in adiabatic models and their
description in terms of 
equation~(\ref{eqn:ultimateequation})
are given in Table~\ref{tab:ultimate}.

In fact, all CDM variants (CDMv)
go through a phase in the radiation-dominated 
epoch when the perturbations in the radiation are pressure 
supported leading
to an essentially smooth component $\Omega_{\rm rad}=\Omega_{S2}$
with $w_2=1/3$ and also
a smooth baryonic component $\Omega_b=\Omega_{S1}$ with $w_1=0$ 
that is held against collapse
by Compton coupling to the photons (see \S \ref{sec:entropygen} 
and \cite{HuSug96}).  
In this case, the clustered
matter is the CDM ($\Omega_{\rm CDM}=\Omega_{C1}$),
and the solution (\ref{eqn:ultimateequation})
describes the evolution of the CDM density perturbations via
the Poisson equation.  The behavior deep in the radiation
domain is given by equation~(\ref{eqn:phismoothregime}),
which says that density perturbations actually grow logarithmically. 
The full solution~(\ref{eqn:ultimateequation}) 
 maps this logarithmic mode into a power-law 
growth across the matter-radiation transition and is useful
for determining the amplitude of small-scale fluctuations 
as a function of the baryon content \cite{HuSug96}.

Smooth components that dominate at late times are also 
described by equation (\ref{eqn:ultimateequation}).  Many
such models have been proposed to reduce the amount of 
small-scale power in the standard CDM model.  The prototypical 
examples are the $\Lambda$CDM model where 
$\Omega_{S2}=\Omega_{\Lambda}= 1-\Omega_m$ ($w_2=-1$)
and the open model
OCDM $\Omega_{S2}=\Omega_{K} = 1-\Omega_m$ ($w_2=-1/3$).  
The former case is special in that
it may alternately be considered a clustered component (see \S 
\ref{sec:stressfreeclustered}).  
The latter case must be considered as a smooth component, and the
growing mode reduces 
to \cite{GroPee75}
\begin{equation}
\Phi \propto {1 \over y} \left(1+ {3 \over y} - {3 \over y }
	\sqrt{1+y \over y} \tanh^{-1} \sqrt{y \over 1+y} \right)\,,
\end{equation}
where $y \propto a$.

\begin{figure}[tb]
\centerline{\epsfxsize=3.6truein \epsffile{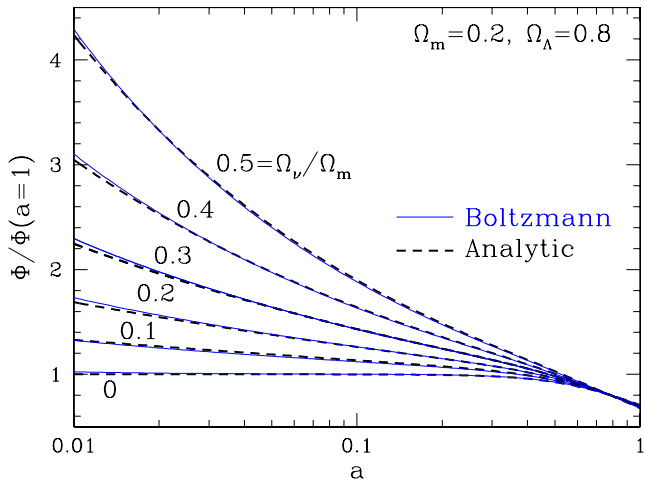}} 
\caption{Two smooth components.  Equation~\protect{(\ref{eqn:ultimateequation})}
is compared with numerical solutions in a $\Lambda$HCDM which has smooth $\Lambda$
and hot (H) dark matter $\Omega_\nu$ components.  Numerical solutions are for modes
well below the Jeans scale of the hot dark matter and discrepancies at early times
are due to radiation contributions in the model.}
\label{fig:lmdm}
\end{figure}

The obvious generalization of such models involves
a smooth component with an equation of state 
that differs from the curvature or
cosmological constant examples.
The prototypical case is the HCDM model where a massive
neutrino (or hot dark matter) component with
$w_1 \approx 0$ remains smooth on small scales
due to residual relativistic effects.  Our general solution 
in fact allows an additional smooth component $w_2$, which
could be curvature (OHCDM), a cosmological constant ($\Lambda$HCDM),
or even some new component with 
an equation of state $w_2=w_Q$ (QHCDM).  
In fact, in the case where there is no hot component, the latter
solution describes the ``quintessence'' model (QCDM) that
has recently received much attention \cite{QChicago,QPaul}.  
Here, a scalar field 
supplies a density component that is smooth inside a sound horizon
that corresponds to the particle horizon.   A comparison of
the analytic solution to numerical results for QCDM and
$\Lambda$HCDM are given in  Fig.~\ref{fig:phis} and \ref{fig:lmdm}.
Likewise, the solution applies to the
GDM generalization \cite{Hu98} where the sound horizon is allowed
to be arbitrary.

Finally, equation~(\ref{eqn:lambdacurvature}) describes the case
of a CDM model with both curvature and cosmological constant 
contributions.  

\begin{table}[h]
\begin{center}
\begin{tabular}{|c|c|c|c|c|l|}
$w_1$ & 
$w_2$ & 
$\Omega_C$ & 
$\Omega_{S1}$ & 
$\Omega_{S2}$ & 
Model 
	\\ \hline
0   			    				 & 
1/3    							 & 
$\Omega_{\rm cdm}$			                 & 
$\Omega_{\rm b}$					 &
$\Omega_{\rm rad}$					 &
CDMv \\
0   			    				 & 
--							 &
$\Omega_{\rm cdm+b}$			                 & 
$\Omega_{\nu}$						 &
-- 							 &
HCDM \\
0   			    				 & 
-1							 &
$\Omega_{\rm cdm+b}$			                 & 
0							 &
$\Omega_{\Lambda}$					 &
$\Lambda$CDM \\
0							 &
-1							 &
$\Omega_{\rm cdm+b}$			                 & 
$\Omega_{\nu}$						 &
$\Omega_{\Lambda}$					 &
$\Lambda$HCDM \\
0   			    				 & 
-1/3							 &
$\Omega_{\rm cdm+b}$			                 & 
0							 &
$\Omega_{K}$						 &
OCDM \\
0							 &
-1/3							 &
$\Omega_{\rm cdm+b}$			                 & 
$\Omega_{\nu}$						 &
$\Omega_{K}$					         &
OHCDM \\
0							 &
-1/3							 &
$\Omega_{\rm cdm+b}$			                 & 
$0$		           				 &
$\Omega_{\rm str}$				         &
strCDM \\
0   			    				 & 
--							 &
$\Omega_{\rm cdm+b}$			                 & 
$\Omega_{\phi}$						 &
--						         &
$\phi$CDM \\
0   			    				 & 
$w_Q$							 &
$\Omega_{\rm cdm+b}$			                 & 
0							 &
$\Omega_{Q}$					         &
QCDM, GDM \\
0   			    				 & 
$w_Q$							 &
$\Omega_{\rm cdm+b}$			                 & 
$\Omega_\nu$						 &
$\Omega_{Q}$					         &
QHCDM, GDM \\
0   			    				 & 
$w_Q$							 &
$\Omega_{\rm cdm+b}$			                 & 
$\Omega_\phi$						 &
$\Omega_{Q}$					         &
$\phi$QCDM, GDM \\
\end{tabular}
\end{center}
\caption{Correspondence between the analytic solution of 
equation (\protect{\ref{eqn:ultimateequation}}) and models with
smooth components. $\Omega_K$ is the fractional effective
density supplied by the curvature component.}
\label{tab:ultimate}
\end{table}	

\subsection{Vector Perturbations}

The behavior of vector modes is far simpler than
that of scalar modes.  
Vector anisotropic stress can be neglected if $H_T^{(\pm 1)} \gg
p\Pi^{(\pm 1)}/\rhoc$ in a gauge where $B^{(\pm 1)}=0$. 
In the absence of stress perturbations, vector modes
simply decay with the expansion (see equation [\ref{eqn:Eulervector}]) 
is solved by
\begin{eqnarray}
(\rho + p) (v^{(\pm 1)} - B^{(\pm 1)}) 
&\propto& a^{-4} \,, \nonumber\\
kB^{(\pm 1)} - \dot H_T^{(\pm 1)} &\propto& a^{-2} \,,
\label{eqn:vectorstressfree}
\end{eqnarray}
The metric source is related algebraically to this quantity by
equation (\ref{eqn:Poissonvector}).   

\begin{figure}[bt]
\centerline{ \epsfxsize = 3.5truein \epsffile{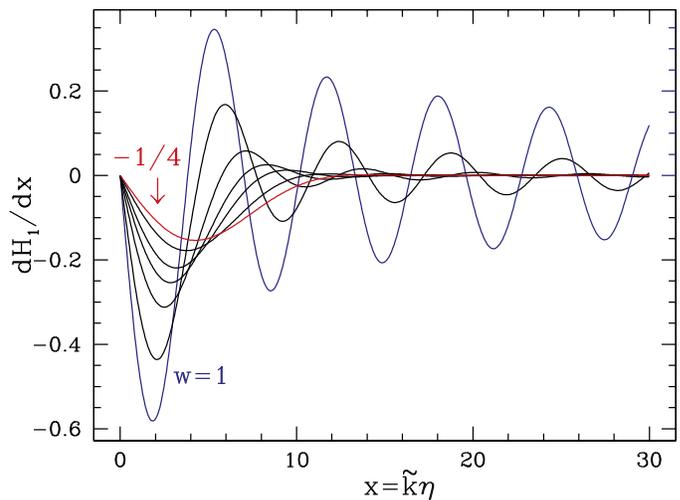}}
\caption{Gravity wave modes.  The amplitude of free gravity waves remains 
constant outside the horizon and oscillates and decays inside in a manner dependent on
the equation of state of the background.  Shown here is the derivative of
the mode since it acts as the source of radiation anisotropies.  Note that
as $w$ increases the oscillatory phase begins sooner relative to horizon 
crossing.}
\label{fig:hdot}
\end{figure}
\subsection{Tensor Perturbations}
\label{sec:tensorfree}

The tensor anisotropic stress is negligible if
$H_T^{(\pm 2)} \gg p\Pi^{(\pm 2)}/\rhocr$.
In this limit,
tensor metric fluctuations
remain constant outside the horizon regardless of
the expansion rate and propagate as free gravity waves inside
of it.  They are described by equation~(\ref{eqn:Poissontensor})
with the sources set to zero.

{\it Full Solutions.---}
If the expansion is dominated by a component with constant
equation of state $\wcr>-1/3$, the fundamental modes of 
gravity waves are
\begin{eqnarray}
H_{1} &=& {2^{m+1}\Gamma(m+3/2)\over\sqrt\pi} x^{-m} j_m (x)\,, \nonumber\\
H_{2} &=& {2^{m+1}\Gamma(m+3/2)\over\sqrt\pi} x^{-m} n_m (x)\,,
\label{eqn:tensorstressfree}
\end{eqnarray}
with $x=k\eta$, $m = (1-3w)/(1+3w)$ and $\tilde k = \sqrt{k^2-2K}$.  Note we
have normalized the modes so that $H_{1}(0)=1$; this mode drops
from unity into damped oscillations 
when the wavelength reaches some fraction of the horizon
that decreases with $m$ and hence increases with $w$. In Fig.~\ref{fig:hdot},
we plot the derivative of these modes, 
as that determines its effect on
the radiation through gravitational redshifts.  

For $w<-1/3$, the universe accelerates and 
the horizon stops growing with the
scale factor.  
This implies that gravity waves will also freeze out at some finite
value related to their amplitude when the universe began accelerating.
The solutions for the gravity wave behavior relative
to the epoch of freeze-out
follow the form of equation~(\ref{eqn:tensorstressfree})
with $x = k \int_\eta^\infty d\eta$.
The $H_{1}$ mode then has damped oscillations
as $x$ increases from negative values but freezes in to 
finite value as $x \rightarrow 0$ 
in the infinite future.

\begin{figure}[bt]
\centerline{ \epsfxsize = 3.6truein \epsffile{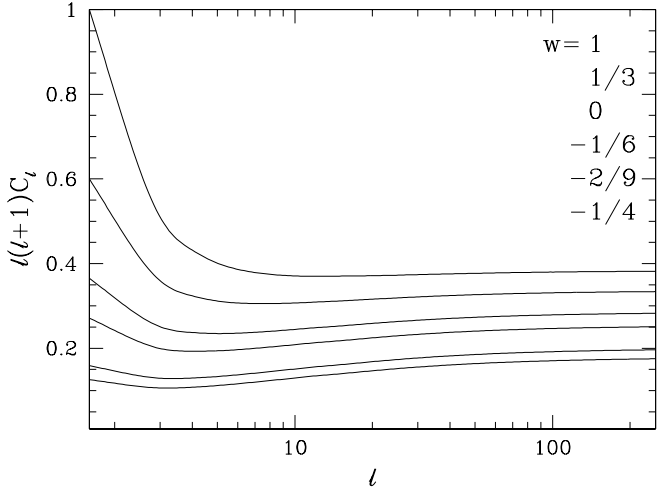}}
\caption{Gravity wave anisotropies generated in free radiation. Numerical solutions
are from a Boltzmann code with scattering sources removed.  Notice that
as $w$ increases so does the anisotropies.  This is directly related
to the behavior of the gravity wave modes in Fig.~\protect{\ref{fig:hdot}}.}
\label{fig:tensorfree}
\end{figure}

{\it Applications.---}
The solutions above help us understand the phenomenology of tensor anisotropies
in the CMB.  The tensor analogue to
equation~(\ref{eqn:thetaell}) is
\begin{equation}
{\Theta_\ell \over 2\ell+1} \approx 
\sqrt{ {3 \over 8} {(\ell+2)! \over (\ell-2)! }} 
\int_0^{\eta_0}d\eta  
e^{-\tau} \dot H_T^{(\pm 2)}
{j_\ell(k\Delta\eta)  \over
	(k\Delta\eta)^2 }\,,
\label{eqn:thetaelltensor}
\end{equation}
where $\Delta\eta=\eta_0-\eta$.
The power spectrum is again given by equation~(\ref{eqn:cl}).
The appearance of the $e^{-\tau}$ damping reflects the fact that
anisotropic stress cannot be supported in the optically thick limit;
we take $\tau \rightarrow 0$ in the examples here to consistently
neglect all anisotropic stress effects.
This in fact is a good
approximation for tensor anisotropies in the 
neutrino background radiation.

For a flat universe with constant $w>-1/3$, the results are shown in 
Fig.~\ref{fig:tensorfree} for the same scale-invariant initial spectrum of
gravity waves, $k^3 |H_T^{(\pm 2)}|^2 = $const.  Notice that
the anisotropies decrease as $w$ is decreased.  Like the scalar ISW effect
discussed in \S \ref{sec:curvtoobs}, the contribution of a decaying tensor mode
to the anisotropy depends on how long the gravity wave takes to decay
relative to the light travel time across the perturbation.
In the limit that it decays before horizon crossing, the photons experience
the full gravitational effect.  If it decays well after horizon crossing,
then the effect suffers cancellation as the photons traverse many wavelengths
of the perturbation.  Thus, the 
relative contribution to the anisotropy 
can be read off the behavior of the normal mode in Fig.~\ref{fig:hdot}.
As $w$ increases, changes in $H_T$ occur at smaller times relative to horizon
crossing.  The anisotropy contribution accordingly goes up.  The effect is
most dramatic for the quadrupole since all $k$-modes above the horizon 
contribute to the quadrupole.
This effect explains why the tensor spectrum in the usual matter-radiation
universe shows an upturn in the spectrum as one goes from modes that
crossed the horizon in the matter dominated epoch to those that crossed
in the radiation dominated epoch (see Fig.~\ref{fig:cltensor}).

For $w<-1/3$ the gravity waves freeze out.  Since CMB anisotropies are driven
by changes in the gravity-wave amplitude, the addition of a $w<-1/3$ component
should suppress anisotropies; this prediction is in agreement with
the effect found in $\Lambda$CDM and QCDM models \cite{Kno95,CalSte98}.

\begin{figure*}[ht]
\centerline{ \epsfxsize = 6.25truein \epsffile{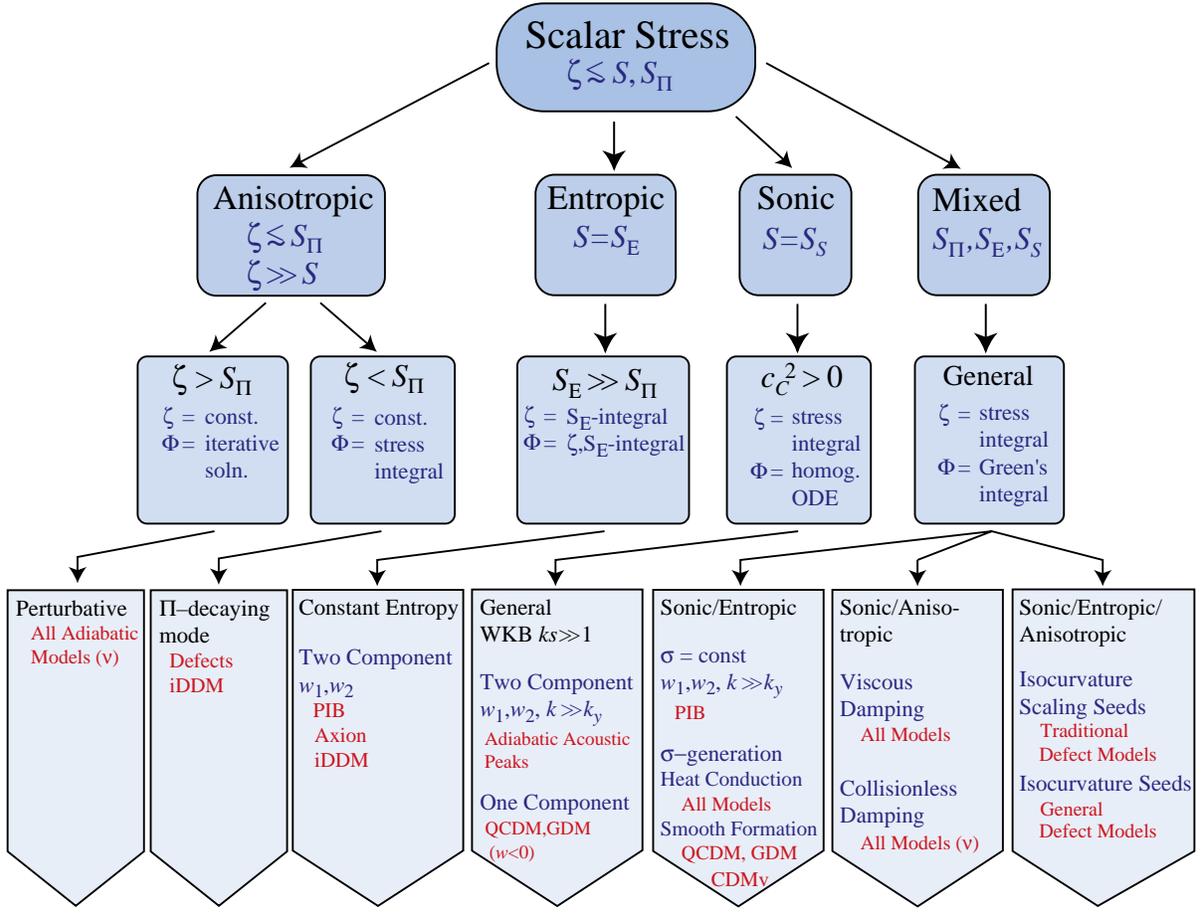}}
\caption{Scalar stress perturbations.  Scalar stress perturbations are divided
into three pure classes: anisotropic, entropic, and sonic.  If one of these is
dominant, then general techniques (third row) can be applied to analyze the
resultant behavior of fluctuations.  Many models (fourth row)
do go through phases where
the stresses are dominated by one of the pure stresses.
The designations ``all models'' and ``all adiabatic models'' 
assume a neutrino background and fluctuations that were present before
last scattering.  Several cases with analytic solutions
illustrate the range of behaviors. }
\label{fig:scalarstress}
\end{figure*}

\section{Pure Stresses}
\label{sec:purestress}

We have shown in the previous section that in certain
regimes stress perturbations can
be ignored.  However,
to provide a complete history of structure formation, one must
track the perturbations across all scales and time.   We will
first consider the pure stress cases in which the dominant
stress contribution is anisotropic, entropic, or sonic,
as defined in \S \ref{sec:stresstypes}.  These prototypical
cases have analytic solutions and are the starting point for
the general cases discussed in \S \ref{sec:mixedstress}.  They also
have direct application in many models.  The anisotropic and sonic
solutions are applicable to all adiabatic
variants of the CDM model (CDMv).
The entropic solutions show how prototypical isocurvature models such
as the baryon (PIB) or axion (AXI) 
isocurvature models form structure. 
We outline these results in Fig.~\ref{fig:scalarstress}.

\subsection{Anisotropic Stress}
\label{sec:anisotropic}

Anisotropic stresses play a special role because they enter directly into the Newtonian metric
through $S_\Pi$, as opposed to other stresses which only
contribute through the causal motion of the matter.   
This is despite the fact that the comoving
curvature $\zetac$ depends only on $\stress$.
In an isocurvature model, only the comoving curvature
need vanish initially, not the Newtonian curvature.

Similarly, 
even though $\zetac$ remains constant above the horizon
in adiabatic models, the Newtonian curvature $\Phi$ evolves under
anisotropic stresses (see Eq.~[\ref{eqn:phiformal}]).
Recall that
\begin{equation}
\stress_\Pi = -8\pi G a^2 \pc \Pic/k^2 = 
	-3\left( {\dot a  \over a} \right)^2 {\pc\over\rhocr}
	 \Pic/k^2 \,,
\end{equation}
such that its effect is enhanced by $(\rhoc'/\rhocr)(k\eta)^{-2}$ 
relative to $\stress$.
Note, however, that once the universe enters a period when
$(\pc/\rhocr) 
\Pic  \ll (k\eta)^2 \zetac$, all traces of the anisotropic
stress from any previous period vanish
in the relation between $\Phi$ and $\zetac$.  Its effect does not 
vanish from the CMB, however, since
anisotropies record a time-integrated history 
of the gravitational potentials.

{\it Full Solutions.\ ---}
To close this system of equations, we need a relation between
$\stressa$ and $\zetac$.  
We will consider two limiting cases: when $\stressa
< \zetac$ as in the case of stresses from radiation backgrounds,
and when $\stressa \gg \zetac$ as is possible with models involving
active sources such as defects.

In the former case, anisotropic stress is 
generally created as a by-product of gravitational instability.
Its anisotropic nature suggests that it can be created from
shear in the velocity and metric, 
and its coordinate transformation properties
demands that its source be gauge invariant.  The linear combination of
these sources that satisfies these requirements is $(kv -H_T)$.
The anisotropic stress can also have a dissipation timescale $\eta_\Pi$.
Together these considerations imply an evolution of the form 
\begin{equation}
\dot \Pic + \eta_\Pi^{-1} \Pic = 4 ( k v - H_T)\alpha\,.
\label{eqn:viscouseomgen}
\end{equation}
In this section, we are interested in large-scale effects and hence
want the longest timescale for the dissipation; this 
is set by the expansion 
time. 
We will consider cases where the timescale is much smaller than the
expansion time in \S \ref{sec:viscous}.

If we take $\eta_\Pi^{-1} = 3\dot a/a$, we recover the phenomenological
parameterization of 
\cite{Hu98}\footnote{$\alpha = c^2_{\rm vis}/w_g$ in the notation of \cite{Hu98}.}
\begin{equation}
\dot \Pic + 3{\dot a \over a} \Pic = 4 ( k v - H_T)\alpha\,,
\label{eqn:viscouseom}
\end{equation}
which has the formal solution
\begin{equation}
\Pic = 4 a^{-3} \int d\eta a^3 (k \vc - H_T)\alpha\, .
\label{eqn:viscousstress}
\end{equation}
From equation~(\ref{eqn:Poissonzeta}), we find
\begin{equation}
\Pic = -{8 \over 3} a^{-3} \int d\eta a^3 \alpha{ k^2 \over {1+\wc}} 
{\rhocr \over \rhoc}
\left( {\dot a \over a} \right)^{-2} \left( \dot \Phi - {\dot a \over a}
\Psi \right) \,,
\end{equation}
for modes well under the curvature scale.
Employing equation~(\ref{eqn:phizetac}) for the zeroth-order potentials
and assuming constant $\alpha$ and $\rhoc=\rhocr$ yields
\begin{equation}
{\Pic \over \zetac}= -2 \alpha (k\eta)^2 { (1+3\wc)^2 \over 
	(4 + 3\wc)(5+3 \wc) } \,,
\end{equation}
for constant $\wc$.
With this source in equation~(\ref{eqn:phiformal}), the Newtonian
potential becomes
\begin{equation}
{\Phi \over \zetac} = {3 + 3 \wc \over 5 + 3 \wc} 
( 1 +  \beta ) \,,
\label{eqn:phizetacpizero}
\end{equation}
with
\begin{equation}
\beta = 16 {\wc \over 1+\wc}{1 \over (4+3 \wc)(5+3\wc)} \alpha \,,
\end{equation}
if the curvature contributes negligibly to the expansion rate.
If $\beta < 1$, 
this process may be repeated
to obtain the desired accuracy.  For example, employing the
second order form for the Newtonian potential
\begin{equation}
{\Psi \over \zetac} 
= -{3+3\wc \over 5+3 \wc} \left\{ 1 - {3 \over 2} \beta [1-{3 \over 2}
	(1+\wc)\beta]
(1+\wc) \right\}\,,
\end{equation}
one can build the third order relation for the curvature
\begin{eqnarray}
{\Phi \over \zetac} &=& {3 + 3 \wc \over 5 + 3 \wc} 
\Big( 1 +  \beta  
\Big\{1-{3 \over 2}
\beta
\nonumber\\
&& \times 
[1-{3 \over 2}(1+\wc)\beta](1+\wc) \Big\} \Big).
\label{eqn:phizetacpi}
\end{eqnarray}

In the opposite limit that
the anisotropic stress contribution is large compared with
the other perturbations 
($\stressa \gg \zetac$, $\stressa \gg \stress$),
the integral solutions for the Newtonian metric reduce to
\begin{eqnarray}
\Phi & = & {\sqrt{\rhoc} \over a} \int {da \over \sqrt{\rhoc}}
\stressa\,, \nonumber\\
\Psi & = & -\Phi + \stressa\,.
\end{eqnarray}
Notice that the integrals remain finite as long as $\stressa$
diverges at zero no faster than $a^{-5/2-3w/2}$ and the 
prefactor $\sqrt{\rho}/a$ is simply the decaying mode
of the Newtonian curvature (see Eq.~[\ref{eqn:phicluster}]).

\begin{figure}[bt]
\centerline{ \epsfxsize = 3.6truein \epsffile{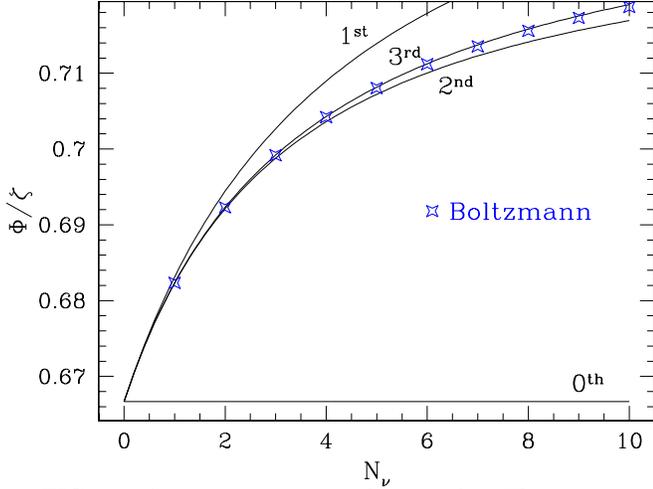}}
\caption{Anisotropic stress at large scales.  The presence of anisotropic stress
affects the Newtonian ($\Phi$) but not comoving ($\zetac$) 
curvature.  If $\stressa < \zetac$, solutions for $\Phi$ 
can be constructed by iteration.  Here we show an example where the
anisotropic stress is produced by massless neutrinos and 
compare numerical results from a Boltzmann code with the iterative solution.
To first order, $\stressa/\zetac =  4 N_\nu/15(4.4 +N_\nu)$,
where $N_\nu$ is the number of massless neutrino species.}
\label{fig:pi}
\end{figure}

{\it Applications.\ ---}
Collisionless radiation provides an application for
these results.  The anisotropic stress of radiation is 
related to the quadrupole moment as defined in 
equation~(\ref{eqn:thetaell}) by
\begin{equation}
\Pi_\gamma = {12 \over 5} \Theta_2 
\end{equation}
on scales much smaller than the curvature radius.  
Equation~(\ref{eqn:thetaell}) also implicitly gives the equation of
motion of the $\ell$th multipole as 
\begin{equation}
\dot\Theta_\ell = k \left[ {\ell \over 2\ell-1} \Theta_{\ell-1} -
			   {\ell+1 \over 2\ell+3} \Theta_{\ell+1} \right]\,.
\label{eqn:hierarchy}
\end{equation}
This is an infinite set of coupled differential equations representing
the fact that radiative stress depends on internal degrees
of freedom of the distribution.  However,
\cite{HuScoSugWhi95} 
introduced an approximation based on the solution of these
equations in the absence of sources
\begin{equation}
{\Theta_\ell \over 2\ell+1} = C j_\ell(k\eta) \,,
\end{equation}
which allows us to express
\begin{equation}
\Theta_{\ell+1} = (2\ell+3)(k\eta)^{-1}\Theta_\ell - {2\ell+3 \over 2\ell-1}\Theta_\ell \,.
\end{equation}
Applying this closure relation to the quadrupole, noting
$(\dot a /a)=1/\eta$ in the radiation-dominated era, and rewriting
the dipole in covariant form 
$\Theta_1 = v_\gamma - H_T$ shows that the anisotropic stress of 
free radiation in the radiation-dominated
era obeys equation (\ref{eqn:viscouseom}) with $\alpha=p_\gamma/\pc 
= \rho_\gamma/\rho_{\rm rad}$.

The photons actually do not behave in this manner before
last scattering since their coupling 
to the baryons destroys any quadrupole moment in the distribution
(see \S \ref{sec:viscous}).
However, the same analysis applies to the massless neutrinos,
whose anisotropic stress can be approximately parameterized by 
$\alpha = \rho_\nu/\rho_{\rm rad}$.
Equation (\ref{eqn:phizetacpi})
shows that the change in $\Phi$ is enhanced
by a factor $1 + (2/15)(\rho_\nu/\rho_{\rm rad})$.  
With equation (\ref{eqn:effectivetemp}), we find the effective
temperature of the CMB to be
\begin{equation}
\Theta_0+\Psi = -{1 \over 2} \Psi = -{1 \over 3}\zetac(0)
\left(1-{4 \over 15} {\rho_\nu \over \rho_{\rm rad}}\right) \,,
\label{eqn:effectivetempnu}
\end{equation}
and the ISW combination to be
\begin{equation}
\Psi-\Phi = -{4 \over 3}\zetac(0) \left(1 - {1 \over 15} {\rho_\nu \over
	\rho_{\rm rad}} \right) \,,
\label{eqn:iswnu}
\end{equation}
to first order. 
These results are equivalent to but more physically
transparent than those of \cite{HuSug95}.  In addition, 
equation~(\ref{eqn:phizetacpi}) introduces second and third order
corrections that are important for models with additional
neutrino species or higher neutrino temperatures $T_\nu$
\begin{equation}
{\rho_\nu \over \rho_\gamma} =  0.681 {N_\nu \over 3} \left( {1.401 T_{\nu} \over
T_{\gamma}} \right)\,,
\end{equation}
as shown in Fig.~\ref{fig:pi}.

Equations (\ref{eqn:effectivetempnu}) and (\ref{eqn:iswnu}) imply that 
the neutrinos have significant but 
opposite effects on the effective temperature and ISW terms
in the CMB anisotropy equation~(\ref{eqn:thetaell}).  The degeneracy
is broken as the fluctuation enters the horizon, as we shall
see in \S \ref{sec:viscous}, leading to detectable effects 
in the acoustic peaks \cite{HuEisTegWhi98}. 

Finally, seed defect sources generally have large anisotropic stress
contributions outside the horizon.  Causality dictates that 
$\pseed \Piseed$ behaves as white noise above the horizon which
together with the so-called scaling ansatz leads to an
$\eta^{-1/2}$ temporal behavior.  This implies that $\stressa
\propto \eta^{-5/2} \propto a^{-5(1+3w)/4}$.  Since
this diverges slower than $a^{-5/2-3w/2}$, it does not 
imply a divergence of any observable \cite{DurKun98} and 
contributes mainly to the decaying mode of the curvature \cite{Bar80}.
Note that these considerations assume vanishing spatial curvature.

\subsection{Entropic Stress}
\label{sec:nonadiabatic}

We next consider the case in which entropic
stress dominates
the other stress components $\stressg \gg \stresssonic, \stressa$.
In this limit $\stressg$ does not depend
on $\zetac$ and equations~(\ref{eqn:zetaformal}) and
(\ref{eqn:phiformal}) are more than
simply a formal solution: they are the integral solutions for 
an arbitrary source in the absence of smooth components.  
For instance, if the entropic stress
has a power-law behavior $\stressg \propto a^n$ then the
comoving curvature 
will have the same behavior 
$\zetac = -\int d\ln a \stressg  \propto a^n$ assuming $\zetac(0)=0$.  

It is 
important to note that once $\stressg$ turns off, $\zetac$ will
remain constant.  Thus entropic stresses that act for only
a fixed amount of time generate curvature fluctuations out
of isocurvature initial conditions that then behave in the
same manner as initial curvature fluctuations.

A natural way of establishing an entropic stress is to have 
two components whose sound speeds 
differ at some point in
the evolution.
It is useful to introduce
the ``entropy''
\begin{equation}
\entropy = 
{\delta\rho_2\over \rho_2+p_2} - {\delta\rho_1\over \rho_1+p_1}\,,
\end{equation}
since its equation of motion is simply
\begin{equation}
\dot\entropy = -k (v_2-v_1)
\label{eqn:entropyode}
\end{equation}
by virtue of equation~(\ref{eqn:continuity}), assuming
no direct energy exchange between species.  
The equivalence
principle guarantees that under purely gravitational evolution
the velocity differences vanish so that a constant $\entropy$ 
is a good approximation outside the horizon.  Tight coupling 
between the components also implies $\dot\entropy=0$.

Let us assume that
the density perturbations are accompanied by sonic
stresses in the rest frame of each component,
\begin{eqnarray}
\Delta\rho_J & = & \delta\rho_J - \dot\rho_J (v_J-v)/k \,,\nonumber\\
\Delta p_J & = & \delta p_J - \dot p_J (v_J-v)/k \,,
\label{eqn:componentsound}
\end{eqnarray}
with $c_J^2 = \Delta p_J/\Delta \rho_J$.   Under the constant
entropy assumption $v_J=v$ and with the definition of the combined
sonic stress
\begin{equation}
c_C^2 = {\rho_1' c_1^2 + \rho_2' c_2^2 \over
		\rho'} \,,
\label{eqn:combinedsound}
\end{equation}
the entropic stress becomes
\begin{equation} 
{\stressg  \over \sigma}
=  \left( \rho_1'\rho_2' \over \rho'' \right)(c_2^2-c_1^2) \,.
\label{eqn:sgammasigma}
\end{equation}
Consequently, entropic stresses are generated when the sound
speeds of the two components differ.  Furthermore
they can be much larger than the sonic stresses if $\delta \rho/\rho 
\ll \sigma$ as is the case for isocurvature models.

{\it Full Solutions. ---} 
If the sonic stresses are also adiabatic ($c_J^2 = p_J'/\rho_J'$),
then $\stressg = \stressna$ and
$\stressna/\sigma = (\rho_1'/\rhoc')'/3$. 
Equation~(\ref{eqn:zetaformal}) then yields
\begin{equation}
\zetac(a) = {\entropy\over 3} {\rho_2' \over \rho'} 
		\Bigg|_0^a  + \zetac(0)\,.
\label{eqn:zetacentropy}
\end{equation}
The Newtonian potential follows from equation~(\ref{eqn:phiformal}).

For example, in the constant $w_1$ and $w_2$ case
\begin{eqnarray}
{\stressna \over \entropy} &=&  {(1+w_1)(1+w_2)\Delta w  \over
	  	[1+w_1+(1+w_2)y]^2 }y \,, \\
{\zetac \over \entropy} &= & {1\over3}{(1+w_2)y\over1+w_1+(1+w_2)y} \,,
\label{eqn:sgamma}
\end{eqnarray}
where recall $\Delta w = w_2-w_1$ and $y=\rho_2/\rho_1$ and we 
have assumed isocurvature initial conditions ($\zetac(0)=0$).
Once $\rho_2$ dominates the energy density, the
entropic stress has been converted into a constant comoving
curvature $\zetac=\entropy/3$.  
This solution may then be substituted into equation
(\ref{eqn:phiformal}) to obtain the integral solution
\begin{eqnarray}
{\Phi \over \entropy} &=& {1\over3}{y\over r+y} + {1\over9\Delta w}{\sqrt{1+y}\over y^n}
\int^y_0 dy{y^{n-1}\over\sqrt{1+y}} \nonumber\\
&& \times
\left[1-{r^2+ry(3\Delta w+1)\over(r+y)^2}\right],
\label{eqn:phientropy}
\end{eqnarray}
where $r = (1+w_1)/(1+w_2)$ and $n=-(5+3w_1)/6\Delta w$.

The Newtonian curvature has the limits
\begin{equation}
{\Phi \over \entropy} 
	= \cases{ \displaystyle{{1+w_2} \over 5 +9 w_1- 6 w_2} y \,,
		& \qquad $(y \ll 1)\,,$ \cr
		\displaystyle{ 1 + w_2 \over 5 + 3 w_2} \,,
		&\qquad $(y \gg 1)$\,.}
\end{equation}
The relation between the Newtonian and comoving curvature in
the $y\gg 1$ limit is exactly the same as the stress-free 
clustered case of \S\ref{sec:stressfreeclustered}.

Another interesting case is when the two components are initially
both radiation $w_1(0)=w_2(0)=1/3$ but the second component
$w_2$ becomes non-relativistic.  
Equation~(\ref{eqn:zetacentropy}) implies that by the time
$\rho_2 \gg \rho_1$, a curvature fluctuation 
\begin{equation}
\zetac = {\entropy \over 3} {\rho_1 \over \rho} \Bigg|_0\,,
\label{eqn:zetacmassive}
\end{equation}
is generated from isocurvature conditions.

The more general solution (\ref{eqn:sgammasigma}) covers
multicomponent models with non-adiabatic sonic stresses
in one or both of the components and for example 
may be applied to scalar field models.

{\it Applications.\ ---}
Entropy perturbations between the matter and the radiation are the
basis of the prototypical isocurvature models, i.e. the baryon 
isocurvature model (PIB) and the axion isocurvature model.
In this case,
the integral in equation~(\ref{eqn:phientropy}) 
can be explicitly solved to obtain \cite{KodSas86}
\begin{equation}
{\Phi\over \entropy} = {16+8y-2y^2+y^3-16\sqrt{1+y}\over 5y^3}\,,
\end{equation}
where $y\propto a$.  Thus, the curvature grows as $a$ in the
radiation-dominated era only to freeze out at the amplitude
reached at matter-radiation equality. 

Motivated by our study, one of us 
constructed an isocurvature
decaying dark matter (iDDM) model that utilizes the mechanism
described above where density fluctuations in
two radiation species are initially balanced \cite{Hu98b}.
The resulting constant superhorizon curvature fluctuation
$\zeta$ is a property generally associated
with adiabatic models.  Another interesting consequence
is that the ratio of large-scale CMB temperature anisotropies 
to the Newtonian potential is neither $1/3$ nor $2$.  This is
because the photons possess initial perturbations.  These 
can be chosen to make the ratio greater
than or less that 2; in particular it can be arranged to be
close to the adiabatic 1/3 relation.

\subsection{Sonic Stress}
\label{sec:adiabatic}

Sonic stresses provide the final pure case.  
Here 
\begin{equation}
\stress = \stresssonic = c_C^2 { \delta \rhoc \over  \rhoc + \pc}\,;
\label{eqn:sonicstress2}
\end{equation}
recall that adiabatic stresses are a special case of
a sonic stress where 
$c_C^2 = c_s^2 \equiv \pc'/\rhoc'$.  
Because the stress fluctuation is related to the comoving density
fluctuation by equation~(\ref{eqn:sonicstress2}) and 
in turn to the Newtonian curvature
via the hybrid Poisson equation (\ref{eqn:hybridpoisson}),
the evolution equation (\ref{eqn:phigeneral})
for the Newtonian curvature 
becomes a homogeneous second order differential equation.
\begin{equation}
{ \rhoc' \over a \rhocr^{1/2} }
\left( { a \rhocr^{1/2} \over \rhoc' }\Phi' \right)'
	+ \left( {1 \over 2} {{\rhocr' +\rhoc'}\over \rhocr} 
	- {\rhoc'' \over \rhoc'} + k^2 s'{}^2 \right)
	\Phi = 0 \,,
\label{eqn:phiadiabatic}
\end{equation}
where the sound horizon is defined as
\begin{equation}
s = \int d\eta\, c_C \,.
\end{equation}
Here, we have assumed that the wavelength is much smaller than the
curvature scale, but all results are applicable to the general
case with the replacement $k \rightarrow \sqrt{k^2-3K}$. 

For $k s' \gg 1$, we can approximate this as an oscillator equation
with an effective mass
\begin{equation}
(m_{\rm eff} \Phi')' + k^2 s'{}^2 m_{\rm eff} \Phi=0\,,
\end{equation}
with
\begin{equation}
m_{\rm eff} =  {a \rhocr^{1/2} \over -\rhoc'} \,.
\end{equation}
Under the WKB assumption that the effective mass is slowly-varying 
compared with the frequency of oscillation, the
fundamental solutions to this equation are
\begin{eqnarray}
\Phi_1  &=& 
	\left({-\rhoc' \over c_C}\right)^{1/2} \cos(ks)\,, \nonumber\\
\Phi_2  &=& 
	\left({-\rhoc' \over c_C}\right)^{1/2} \sin(ks)\,. 
\label{eqn:oscasymptotic}
\end{eqnarray}
This equation says that
once the fluctuation passes inside the sound horizon, the Newtonian
potential oscillates and decays reflecting analogous behavior
in the density perturbation through the Poisson equation.  

{\it Full Solutions.\ ---}
Equation~(\ref{eqn:oscasymptotic}) suggests
that a change of variables to
\begin{equation}
Q = \left({c_C \over -\rhoc'}\right)^{1/2}\Phi \,,
\end{equation}
should simplify the equations of motion.  
Indeed, equation (\ref{eqn:phiadiabatic}) becomes
\begin{equation}
{d^2 Q\over d(ks)^2} + \left(1-{F\over k^2s^2}\right)Q = 0 \,,
\label{eqn:Qbessel}
\end{equation}
where 
\begin{eqnarray}
\label{eqn:QbesselF}
F &=& -{s^2\over 2s'^2}\left\{
{\rhocr'+\rhoc'\over\rhocr}-{\rhoc''\over\rhoc'}+{\rhoc'''\over\rhoc'}
-{3\rhoc''^2\over 2\rhoc'^2}
\right.\\ & &\left.
+{\rhocr'\over2\rhocr}\left({\rhoc''\over\rhoc'}-{c_C'\over c_C}\right)
-{c_C'\over c_C} -{c_C''\over c_C} + {3c_C'^2\over4c_C^2}\right\}\,.
\nonumber
\end{eqnarray}
If $F$ is constant, then equation (\ref{eqn:Qbessel}) is a variant of
the Bessel equation, and the solutions $Q$ will approach $\sin(ks)$
(up to a phase) at large $ks$.  Furthermore, when $ks\gg |F|^{1/2}$, 
the term with $F$ may simply be neglected, and again the solutions will
be $\sin(ks)$ with arbitrary phase.
Because the latter solutions will match trivially onto the former, 
we reach the conclusion that if, for a given mode, $F$ is constant until
$s\gg |F|^{1/2}/k$, 
then the appropriate Bessel function solution will hold
for {\it all} times, regardless of how $F$ varies once 
$ks\gg |F|^{1/2}$.
If $c_C$ and $\rhoc$ vary on the Hubble time scale, then
$F$ is order unity, and the solution describes modes that are well inside
the horizon before $F$ begins to vary.

The simplest way to arrange $F$ to be constant is to have a constant
equation of state $w_1$ and constant sound speed $c_C$.  For constant $w_1$,
$s$ is only defined for $w_1>-1/3$, and we therefore set $\rhoc=\rhocr$.
Then
\begin{equation}
F = 6{1+w_1\over(1+3w_1)^2},
\end{equation}
independent of the sound speed.
The growing and decaying solutions for $\Phi$ are then
\begin{eqnarray}
\Phi_1  &\propto& 
	\left({-\rhoc' ks\over c_C}\right)^{1/2} J_\nu(ks)
	\,, \nonumber\\
\Phi_2  &\propto& 
	\left({-\rhoc' ks \over c_C}\right)^{1/2} N_\nu(ks)
	\,, 
\label{eqn:phiadiabaticsol}
\end{eqnarray}
where $\nu =(5+3w_1)/(2+6 w_1)$.   
Again, once a given mode is well
inside the sound horizon, the condition that $w_1$ and $c_C$ 
be constant
can be relaxed.


{\it Applications.\ ---}
Adiabatic stress perturbations in
a baryon-photon universe provide the prototypical example.  
In this case, the tight coupling between the photons and baryons
through Compton scattering prevents the generation of entropy
through the motion of matter
(see \S \ref{sec:entropygen}). 
The growing mode of equation (\ref{eqn:phiadiabaticsol}) becomes
\cite{KodSas86}
\begin{equation}
{\Phi\over \zetac(0)} = {3\over y^2} 
	\left({k_{y} \over k}\right)^2 (1+{3 \over 4}y)^{3/4}
	\left[ {\sin(ks) \over ks} - \cos(ks) \right]\,, 
\label{eqn:phiacousticadi}
\end{equation}
with $k_y = (\dot a/a)_{y=1}$ and
\begin{equation}
s = {4\sqrt{2} \over 3} {1 \over k_{y}} 
	\ln{ \sqrt{4+3y}  + \sqrt{3y+3} \over 2 + \sqrt{3} } \,.
\end{equation}
We have chosen the normalization to match the stress-free solution
$\Phi = 2\zetac/3$ initially.
In this case, the acoustic oscillations of the photons are directly
related to behavior of the potential via the Poisson equation
\begin{eqnarray}
\Theta_0 &=& {1 \over 3} \left( {k \over k_{y}} \right)^2 {y^2 \over
1+3y/4} \Phi \label{eqn:theta0}\nonumber\\
       &=&  (1+3y/4)^{-1/4}\zetac(0) \cos(ks) \qquad (ks \gg 1)\,.
\end{eqnarray}
Since the acoustic oscillations are responsible for the acoustic
peaks in the CMB from equation~(\ref{eqn:thetaell}), this determines
the morphology of the acoustic peaks in this simple adiabatic 
photon-baryon universe.  
The important result is that the acoustic oscillations follow a
$\cos(ks)$ pattern in phase with an amplitude 
that is enhanced by the decay of the initial curvature perturbation.
These results are in fact generic to adiabatic models due to 
similar evolution in the driving potentials \cite{HuWhi96}. 

\begin{figure}[bt]
\centerline{ \epsfxsize = 3.6truein \epsffile{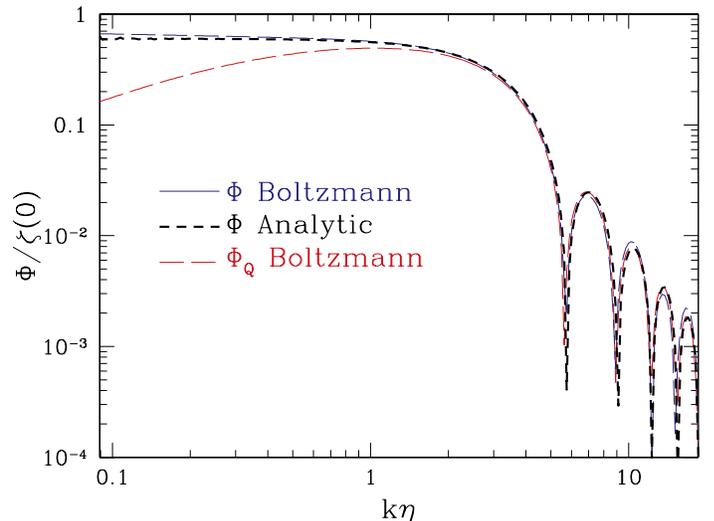}}
\caption{Sonic stresses in a scalar-field dominated universe.  Sonic stresses cause
scalar-field perturbations to stop growing and hence lead to a decay in the 
Newtonian curvature in a scalar-field dominated universe.  Here we compare
a QCDM model with $w_{Q}=0$ and no CDM with the 
analytic prediction of 
equation~(\protect{\ref{eqn:phiadiabaticsol}}).  The discrepancy at early
times is due to radiation contributions in the QCDM model and those at late times
from the baryons.  }
\label{fig:xosc}
\end{figure}

A second example is provided by the QCDM and GDM models where
the equation of state can go negative $(w<0)$ while the 
comoving sound speed remains real $(c_C^2>0)$.  
In QCDM, 
$c_C^2 \equiv 1$ by virtue of the scalar-field equations of motion.
In GDM, it is allowed to have any positive value.  The
solutions are strictly valid only after the exotic component 
dominates the expansion rate and fluctuations.  To test this
solution, we show an extreme example where the dark 
matter is all in a ``Q'' component
(no CDM) with $w_Q=0$ (see Fig.~\ref{fig:xosc}). The small departures
at early and late times are due to the radiation and baryonic 
contributions. 

\section{Mixed Stresses}
\label{sec:mixedstress}

Although the purely anisotropic, entropic and sonic stress cases 
of the last section illustrate
many aspects of stress phenomenology, 
when all three are present they can interact and
create a diverse range of behavior.  
In this section, we study a few typical cases: sonic
stress generation from entropic stress, entropic stress generation
from sonic stress, and anisotropic stress dissipation of sonic
stress.   The first process is responsible for generating 
acoustic phenomena in all isocurvature models.  The second process
is responsible for the heat conduction in fluids and the generation 
of smooth components.  The last process is responsible for 
viscous dissipation of acoustic phenomena and also provides 
an alternate means of generating a smooth component.  
We briefly discuss seed (defect) stresses, which carry not
only all three types of scalar stress but also vector and
tensor stresses as well.   Finally, we consider the effect
of tensor anisotropic stress on gravitational waves and CMB
anisotropies in passive models.

\subsection{Formal Solution}

If the equations of motion for a sonic stress perturbation
are known to be solved by $\Phi_1$ and $\Phi_2$, the full solution
can be written in integral form as,
\begin{eqnarray}
\label{eqn:greensphi}
\Phi& = & A \Phi_1(a) + B\Phi_2(a) \\
	&& + \int_0^a {d \tilde a \over \tilde a}
	{ \Phi_1(\tilde a) \Phi_2(a) - \Phi_1(a)\Phi_2(\tilde a) 
	\over \Phi_1(\tilde a) \Phi_2'(\tilde a)    
	- \Phi_1'(\tilde a) \Phi_2(\tilde a)  }    \nonumber\\
	&&\times \left[ {1 \over 2} {\rhoc' \over \rhocr}
	\stressg
 	+ \stressa' + \left( {1 \over 2}{ \rhocr'+\rhoc' \over\rhocr }
	- {\rhoc'' \over \rhoc'} \right) \stressa\right]\,, \nonumber
\end{eqnarray} 
where $A$ and $B$ are arbitrary constants.
This solution is merely formal unless the remaining stress sources 
can be specified independently of $\Phi$.   We now consider 
some special cases.

\subsection{Entropic/Sonic Stress}

A model that begins with isocurvature initial conditions will 
generate adiabatic density fluctuations through the action of
entropic stresses, as discussed in \S\ref{sec:nonadiabatic}.  These density perturbations carry with
them sonic perturbations through equation 
(\ref{eqn:sonicstress})
that stop the further growth of density perturbations through the
same pressure support mechanism discussed in \S\ref{sec:adiabatic}.
Therefore, entropic growth of curvature
fluctuations generally ceases once
the
fluctuations cross the sound horizon of the dominant species.

{\it Full Solutions.\ ---}
The solutions for $\Phi$ in equation~(\ref{eqn:phiadiabaticsol}) 
along with the entropic stress from equation
(\ref{eqn:sgammasigma}) in equation~(\ref{eqn:greensphi})
allow us to
construct the full solution in the presence of a constant entropy 
and $\rho=\rhocr$
\begin{eqnarray}
{\Phi \over \entropy} &=& -{2 \over 3}\pi^2 G 
	\left( {-\rhoc' s \over c_C} \right)^{1/2}
	\int d\tilde\eta Y(\tilde\eta) 
        \nonumber\\
	&&\times \left[ J_\nu(k\tilde s) N_\nu(k s) - J_\nu(k s)
        N_\nu(k\tilde s)\right] \,,
\end{eqnarray}
where
\begin{eqnarray}
Y(\eta) = a^2 \left( {-\rhoc' s \over c_C} \right)^{1/2} 
	\left( \rho_1' \rho_2' \over \rho'' \right)(c_2^2-c_1^2) \,,
\end{eqnarray}
Recall that the combined and component sound speeds 
were defined in equations~(\ref{eqn:combinedsound}) and
(\ref{eqn:componentsound}) respectively.
This solution is valid for $k/k_{y} \gg 1$ and $k \gg |K|^{1/2}$
(see \S \ref{sec:adiabatic}).

{\it Applications.\ ---}
The case of $w_1=1/3$ and $w_2=0$ is of special interest because it corresponds
to baryon-photon entropy perturbations, as in the PIB model before
last scattering.
The result
is that on small scales where
$ks \gg 1$ and $k/k_{y} \gg 1$, the curvature behaves 
as \cite{KodSas86}
\begin{equation}
{\Phi\over\entropy} = {3 \over 4y} \left({ k_{y} \over k}\right)^2
 \left[ 1 - {\sqrt{3} \over 2}
{k_{y} \over k} {(4+ 3y)^{3/4} \over y} \sin(ks)\right],
\end{equation}
which is again directly related to the temperature oscillations by
equation~(\ref{eqn:theta0}).  The first term in brackets is due
to density perturbations in the baryons remaining from the
constant entropy condition $\sigma = \delta_b - {3 \over 4}\delta_\gamma
\approx \delta_b$.  
The second term represents decaying acoustic
oscillations from the adiabatic pressure.
The extra factor of $k_y/k$ reflects the fact that the curvature
grows as $a$ until sound horizon crossing at $a_H \propto (k_y/k)$
\cite{PreVis80}.

An interesting result is that
the acoustic oscillation follows a $\sin(ks)$ relation, implying 
the opposite phase in the acoustic peaks compared with the $\cos(ks)$
adiabatic case \cite{HuSug95}.   
This result is rather generic to isocurvature
models again due to the similar behavior of the driving potentials.
For example, axionic isocurvature models
where the entropy is between the radiation and the CDM also 
follows this pattern.  

For this reason, isocurvature seed pressure also tends 
to generate this type of
acoustic pattern \cite{HuWhi96}.  These stresses are found 
in topological defect models; indeed the dominant scalar
mode of strings, monopoles, textures, and non-topological
textures do behave in this manner \cite{PenSelTur97}.
However, only in the latter two are the other modes 
sufficiently small at the first few peaks 
to yield clean acoustic peaks even 
for the scalar perturbations alone.  
Defect models generically 
have vector and tensor stresses that generate comparable levels
of anisotropy and further obscure acoustic phenomena.

It is possible to construct isocurvature models
with an ``adiabatic'' pattern of acoustic peaks.  
The simplest way to arrange this is to create constant
comoving curvature perturbations $\zeta$ through entropic
stresses that turn off well before the perturbation crosses
the horizon.  A concrete example of this kind of mechanism
is given in \cite{Hu98b}.  As is clear from \S \ref{sec:nonadiabatic}
and originally pointed out by \cite{HuWhi96},
for an isocurvature  model to generate {\it scale-invariant} 
curvature perturbations requires that the entropic stress
have superhorizon scale correlations which cannot be generated
causally without an inflationary epoch. 

\subsection{Sonic/Entropic Stress}
\label{sec:entropygen}

Likewise, an adiabatic or 
sonic fluctuation will not generally remain so as 
it crosses the horizon.  
Inside the horizon, the fact that components with different equations of
state have different pressure responses to gravitational compression
will cause the species to move independently.  
The generation of entropic stresses is in fact
a primary mechanism for creating the
smooth components of matter discussed in \S \ref{sec:stressfreegeneral}.

{\it Full Solutions.---}
The case of two components with constant $w_1$ and $w_2$ again provides
an instructive example.  The equation of motion for the entropy 
(\ref{eqn:entropyode}) is constructed
out of combining the continuity equations (\ref{eqn:continuity}) 
of the two species and has the 
formal solution
\begin{equation}
\entropy =  \entropy(0) -k \int d \eta (v_2 - v_1) \,.
\label{eqn:entropygeneration}
\end{equation}
Recall the entropy is related to the entropic stress by 
equation~(\ref{eqn:sgammasigma}). Note that entropy leads to 
entropic stress only if the two components differ in their
equation of states ($\Delta w \ne 0$).

Since we are interested in the generation of entropy, we will assume that
the initial entropy perturbation $\entropy(0)$ vanishes.  Two interesting
cases are when the entropy generation timescale $\eta_\Gamma$ is small compared to the
expansion time and when it is comparable to the expansion timescale.

In the former case, the entropy will lead to dissipative behavior if
\begin{equation}
\entropy \approx -k\eta_\Gamma \vc \,,
\label{eqn:entropyassumption}
\end{equation}
where $\eta_\Gamma$ is some characteristic timescale for entropy generation.
While this form is not gauge invariant even though $\entropy$ is,
the ambiguity vanishes inside the horizon.  
Hence, equation~(\ref{eqn:entropyassumption}) is 
a good approximation in the desired case $\eta_\Gamma \ll \eta$.  
Substituting equation~(\ref{eqn:entropyassumption}) into
(\ref{eqn:sgammasigma}) and then (\ref{eqn:Euler}) in 
Newtonian gauge and assuming
a solution of the form $v \propto \exp(i\int \omega d\eta)$,
we obtain the dispersion relation
\begin{equation}
\omega = \pm k c_C - i {1 \over 2} {\rho_1'\rho_2' \over \rho'^2} 
(c_2^2-c_1^2) k^2 \eta_\Gamma \,,
\label{eqn:heatdispersiongeneral}
\end{equation}
where recall $c_C$ was defined in equation~(\ref{eqn:combinedsound}).
Note that we have assumed $\eta_\Gamma/\eta \ll 1$ in order 
to replace rest-frame sound speeds with comoving sound 
speeds.

This describes an exponential damping of sound waves as the wavelength passes
the ``diffusion'' scale $k = (\eta_\Gamma \eta)^{-1}$.

In the opposite regime, where the entropy is generated on the horizon scale,
the sonic nature of the total system breaks down before acoustic oscillations
even start.  In this case, components can decouple completely and form subsystems
where sonic, anisotropic, or smooth effects can occur separately.
Here it is simpler to describe the behavior of individual
subsystems through variables that are comoving 
with respect to the individual species $J$.   
The sound speed $c_J$ and the contribution to the Newtonian
potential are useful (see Eq.~[\ref{eqn:componentsound}])
\begin{eqnarray}
\Phi_J & = & 4\pi G \Delta \rho_J/(k^2-3K) \,,
\end{eqnarray}
which implies $\Phi  =  \sum_J \Phi_J$.
If the density fluctuations in the other species are still below 
their own sound horizon, then the oscillating components can become
effectively smooth in comparison.  From that time forward, we can
treat the system as having a smooth component, and the curvature
fluctuation contributed by the other species is governed by
equation~(\ref{eqn:phismoothclustered}).

\begin{figure}[bt]
\centerline{ \epsfxsize = 3.6truein \epsffile{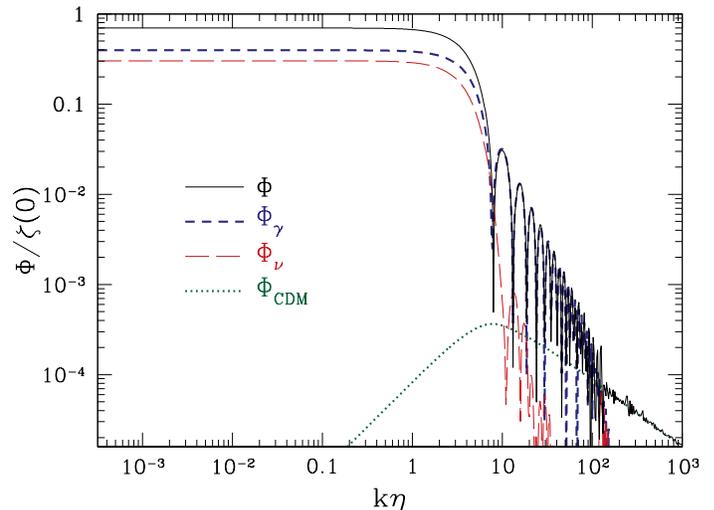}}
\caption{Creation of a smooth radiation background.  Numerical results in a $\Lambda$CDM
model are shown here.  Notice that upon crossing the sound horizon, the photon contribution
$\Phi_\gamma$
to the total curvature $\Phi$ damps and oscillates, while the neutrino
contributions damp much more rapidly due to collisionless damping from its anisotropic
stress.  The oscillating photon perturbations yield little time-averaged effect,
and the CDM evolves under equation~(\protect\ref{eqn:ultimateequation}) leading to the well-known
logarithmic tail in the CDM transfer function \protect\cite{BarBonKaiSza86}.} 
\label{fig:matrad}
\end{figure}

{\it Applications.\ ---}
The case of a short entropic timescale $\eta_\Gamma \ll \eta$ is realized in
the photon-baryon fluid before recombination and is relevant for considering
the damping of acoustic phenomena in the CMB for all models of structure 
formation \cite{Sil68,PeeYu70}.
Here the entropic time scale is derived from the baryon Euler
equation
\begin{eqnarray}
v_b-v_\gamma & \approx & \dot \tau^{-1} {R} \dot v \,,
\end{eqnarray} 
where $R=3\rho_b/4\rho_\gamma$.
Inserting this into equation~(\ref{eqn:entropygeneration}), we obtain
$\eta_\Gamma = \dot\tau^{-1} R$ under the rapid oscillation assumption.
The dispersion relation then becomes \cite{PeeYu70}
\begin{equation}
\omega = \pm k c_s + i{ 1\over 6} k^2 \dot\tau^{-1} {R^2 \over (1+R)^2} \,.
\label{eqn:heatdispersion}
\end{equation}
These entropic pressure techniques provide a simpler and
more transparent derivation
of this well-known result than exists in the literature.
Note that heat-conduction damping is suppressed by $R$ in the
photon-dominated epoch, and we shall see that in that case viscous damping
from the anisotropic stress is more important.

The opposite limit of an entropic timescale on order the expansion time
$\eta_\Gamma \sim \eta$ is applicable to all models with CDM.  The CDM
never participates in the acoustic oscillations of the baryon-photon
system even during the radiation-dominated era.
In Fig.~\ref{fig:matrad}, we show
an example of a perturbation deep in the radiation-dominated era
in a CDM model with the usual neutrino content.
Before horizon crossing $k\eta \ll 1$, the perturbations are
adiabatic and the total potential $\Phi$ is constant with contributions
from the photons $\Phi_\gamma$ and the neutrinos $\Phi_\nu$.  
The sound horizon for the radiation is $s = \eta/\sqrt{3}$ and
once $k\eta \gg \sqrt{3}$, the potential contribution of the
photons starts to decay as in equation~(\ref{eqn:phiacousticadi}).
The neutrino contributions decay even faster due to their
anisotropic stress, as we will discuss in the next section. 
The CDM contribution $\Phi_{\rm CDM}$ then turns over and behaves
as $\Phi_{\rm CDM} \propto \ln(C a)/a$ where $C$ is some constant
\cite{HuSug96}.  This is exactly the behavior predicted by 
equation~(\ref{eqn:phismoothclustered}) assuming the 
radiation component is
smooth.  In actuality, the photon component is not smooth compared
with the CDM component but oscillates sufficiently rapidly that its
time-average is negligible. 

\begin{figure}[bt]
\centerline{ \epsfxsize = 3.6truein \epsffile{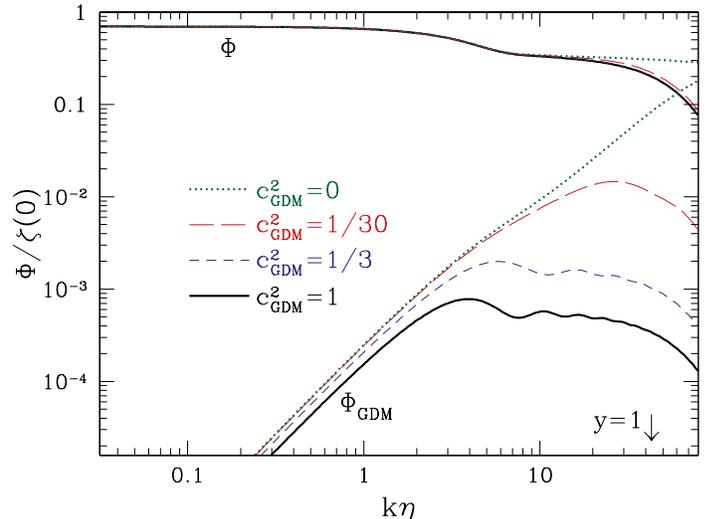}}
\caption{Creation of a smooth quintessence or GDM component with $w_{\rm GDM}=-1/3$.  
As shown in Fig.~\ref{fig:xosc}, sonic stresses in the GDM component prevent perturbation
growth inside the sound horizon $c_{\rm GDM} k\eta \sim 1$.  If this occurs well before
GDM domination at $y=1$, then the total Newtonian curvature will evolve as if
the component were always smooth (i.e. under 
equation~[\protect{\ref{eqn:ultimateequation}}]).  If this occurs well after, the Newtonian
curvature will behave as if all components were fully clustered
(i.e. under equation~[\protect{\ref{eqn:phiclusteredsoln}}]).  For a given scale,
this depends on the comoving sound speed $c_{\rm GDM}$; for quintessence
$c_{\rm GDM}=1$.}
\label{fig:scalarfield}
\end{figure}

These considerations also apply to models in which 
an additional component with
$w_{\rm GDM} < 0$ comes to 
dominate at late times.
Quintessence and 
GDM models (see Fig.~\ref{fig:scalarfield}) are examples thereof. 
In \S \ref{sec:adiabatic}, we considered the case where the
GDM completely dominated the expansion.  Here we consider
the effect of adding a component of CDM.
 
The critical parameter here is the comoving
sound speed of the GDM $c_{\rm GDM}$.  Slowly-roling scalar fields
found in the QCDM subcategory have $c_{\rm GDM}=1$, but components
with lower sound speeds are possible in principle.  The sound speed tells us
how long after horizon crossing the GDM component stabilizes due to
pressure support.  If the perturbation is already within the sound
horizon by the time the GDM comes to dominate the expansion rate $(y=1)$,
then the total potential will behave under 
equation~(\ref{eqn:phismoothclustered}) as if the potential is smooth
regardless of the exact value of the sound speed.  
As the sound speed is lowered such that 
crossing occurs near $(y=1)$, we
see effects from the finite sound speed.  In the limit where
the perturbation remains above the sound horizon until the present, 
the solution returns to the clustered case of equation~(\ref{eqn:phicluster}).

\subsection{Sonic/Anisotropic}
\label{sec:viscous}

We have seen in \S \ref{sec:anisotropic} how anisotropic stress
can generate Newtonian curvature perturbations
but, like adiabatic stress, it can also destroy them.   
In this context, anisotropic stress represents the
``frictional force'' set up in response to the non-uniform bulk
flow of matter and shear in the metric.   
As we shall see, this sort of behavior is not confined to fluids.
It represents another mechanism for generating
a smooth component.

{\it Full Solutions.---}
The phenomenological parameterization of anisotropic stress in
equation~(\ref{eqn:viscouseomgen}) yields two interesting limits.
The short timescale limit ($k\eta_\Pi \ll 1$, $\eta_\Pi/\eta \ll 1$)
leads to viscous or collisional damping.
Here, the anisotropic stress is algebraically related
to the velocity in shear-free frames 
\begin{equation}
\Pic = 4\alpha \eta_\Pi (k \vc - H_T)\,.
\label{eqn:pidiffusive}
\end{equation}
Under the same assumptions used to derive the dispersion relation
for heat conduction (\ref{eqn:heatdispersiongeneral}),  
we find the viscous dispersion relation 
\begin{equation}
\omega = \pm k c_C + i {4\over 3} 
	k^2\alpha \eta_\Pi {\pc \over \pc+\rhoc} \alpha\,,
\end{equation}
implying dissipation at a characteristic 
scale $k \sim 1/\sqrt{\eta\eta_\Pi}$.

If the dissipation timescale is comparable to the expansion timescale,
the damping occurs at horizon crossing but is more gradual.  
The formal solution in this limit is given in equation~(\ref{eqn:viscousstress})
and approximates the effects of collisionless damping.

{\it Applications.\ ---}
Collisional damping occurs in the photon-baryon fluid before last 
scattering and is the primary dissipation mechanism for acoustic
oscillations the CMB \cite{Sil68}. 
Equation~(\ref{eqn:pidiffusive})
then describes the anisotropic stress of the photons $\Pi_\gamma$ 
with  $\alpha = 2/5$ and $\eta_\Pi ={4 \over 3} \dot\tau^{-1}$.
Repeating the calculation leading to 
equation~(\ref{eqn:heatdispersion}), the dispersion relation
for the oscillations becomes \cite{Sil68,Kai83}
\begin{equation}
\omega = \pm k c_C + i {8 \over 45} 
k^2 \dot\tau^{-1} 
{1 \over {1+R}}\,.
\label{eqn:viscousdispersion}
\end{equation}
In comparison to the heat conduction dissipation of equation~(\ref{eqn:heatdispersion}),
viscous dissipation is more effective if $R<1$, which is the case for the
baryon content implied by big bang nucleosynthesis 
(e.g. \cite{SchTur98}). 

Collisionless damping occurs in free radiation. 
Free radiation behaves in this ``frictional'' manner because 
gradients in the potential flow are
dissipated as radiation streams from one part of the flow to the next.
In terms of the multipole moments, power in the dipole gets transferred
to the quadrupole and so on through the hierarchy equations 
(\ref{eqn:hierarchy}).  As such, the anisotropic stress acts as 
the gateway for anisotropy generation in the CMB.  It also 
leads to a damping of density and velocity perturbations inside the horizon 
for all species of free radiation and is
the reason that in Fig.~\ref{fig:matrad} the neutrino contributions
damp more rapidly than the photon contributions.  
More generally, it is responsible for smoothing out components
when entropic stresses cannot be generated, e.g. when $c_2^2-c_1^2=0$. 

These dissipational terms may also be important in stabilizing
other forms of matter.
One way to generate a smooth density component
is to introduce an effective viscous rather than sonic
stress \cite{Hu98}; a mechanism of this type 
is thought to be involved in stabilizing
the strCDM model \cite{SpePen97}.
Turok \cite{Tur97} suggested an extreme example of this sort, in which
the comoving sound speed of the seeds is imaginary but the anisotropic stress
is perfectly balanced to counter the otherwise exponential growth of density
fluctuations. 

\subsection{Sonic/Entropic/Anisotropic Seed Stress}
\label{sec:sea}

The seed stresses of topological defect models provide an example
where all types of stresses coexist.  
While the consequences of a given seed stress for structure formation
are straightforward to work out, the behavior of the seed stress
itself is more difficult and requires simulations
to work out in detail.  

{\it Full Solutions.---}
The
two-point statistics of
complicated defect models can be accurately modeled as the incoherent (quadrature) 
sum of a relatively small number ($\sim$ 10-20) of  simple seed 
stress histories   
\cite{Tur97,PenSelTur97}.
Each individual source
may then be determined by the techniques above (see Eq.~[\ref{eqn:continuityseed}] and
[\ref{eqn:Eulerseed}]).

{\it Applications.---}
The simplest defect models typically have two other properties that have important
phenomenological consequences.  Defect models are causal in the classical
sense.  The stress perturbations $\delta\pseed$ 
and $\pseed\Piseed$ must fall off at least as white noise $(k^{0})$ 
outside the horizon and  the initial curvature must
vanish $(\zetac(0)=0)$.  The traditional string and texture models also
obey a scaling relation that states that the stress histories depend on 
wavelength only through the combination $k\eta$.  This ensures self-similarity
of the structure at horizon crossing and leads to nearly 
scale-invariant
CMB anisotropies.  Because the simplest versions of these models run into
difficulties when CMB anisotropies are compared with large-scale structure,
phenomenological models that do not obey the scaling relation have 
recently received some attention \cite{AlbBatRob97b}.

\subsection{Vector Stress}
\label{sec:vectorstress}

We next consider the effect of vector 
anisotropic stresses.  Recall that in the absence of vector stress,
the vector perturbation decays.  In order to generate
an observable effect, vector perturbations must be continuously generated
by vector anisotropic stresses.   This implies that vector modes are
unique to the active stress models, of which defects are
an example.

{\it Full Solutions.---}
The solutions in the presence of stress sources can be 
constructed via Green's function techniques from the stress-free
solutions of (\ref{eqn:vectorstressfree}).
For the vector modes, the solution becomes
\begin{eqnarray}
v^{(\pm 1)} - B^{(\pm 1)}
&=& a^{-4}(\rho+p)^{-1}
	\Big[ C -{1\over 2}{(1-2K/k^2)}\nonumber\\
	&&\quad\times k\int_0^\eta d\tilde\eta
	a^4 p \Pi^{(\pm 1)}\Big] \,,
\label{eqn:vectorstresssoln}
\end{eqnarray}
where $C=$const.\ represents the decaying mode.
The remaining metric perturbation $k B^{(\pm 1)} - \dot H_T^{(\pm 1)}$
is related algebraically
to equation~(\ref{eqn:vectorstresssoln}) through
equation~(\ref{eqn:Poissonvector}).

{\it Applications.---}
The discussion
of scalar seed stress in defect models 
in \S \ref{sec:sea} also apply to vector stresses with a few additional
considerations.  
Defect models generally have comparable scalar, vector, and tensor 
anisotropic stress sources above the horizon \cite{TurPenSel98}.  
Since CMB anisotropies
are primarily generated at horizon crossing, 
these sources tend to yield comparable
anisotropy contributions for modes that
cross after last scattering.  Vector 
modes that cross before last scattering
do not contribute due to suppression of tensor anisotropies in the
CMB from scattering.  Thus, vector modes can obscure the first few
scalar acoustic features in defect models.   On the other hand,
vector modes have a special signature in the CMB polarization that
may assist in their isolation \cite{HuWhi97,SelPenTur97}.  

\subsection{Tensor Stresses}
\label{sec:tensorstress}

Tensor anisotropic 
stresses provide sources and sinks for gravity waves.  
It is well known that quadrupolar stresses generate gravity
waves.  Furthermore, a passing
gravity wave will also impart some energy to the radiation 
backgrounds via differential gravitational redshifts and thereby decay.

{\it Full Solutions.---}
An integral solution may be 
constructed out of the homogeneous solutions of \S \ref{sec:tensorfree}
as
\begin{eqnarray}
\label{eqn:tensorgreens}
&&H_T^{(\pm 2)}(\eta) =  C_1 H_1(\eta) + C_2 H_2(\eta)  \\
&&\quad+ \int_0^\eta d \tilde\eta
	{ H_1(\tilde \eta) H_2(\eta) - H_1(\eta)H_2(\tilde \eta) 
	\over H_1(\tilde\eta) H_2'(\tilde \eta)    
	- H_1'(\tilde \eta) H_2(\tilde \eta)  }    
	8\pi G a^2 p\Pi^{(\pm 2)} \nonumber,
\end{eqnarray} 
where $C_1$ and $C_2$ are constants associated with the initial 
conditions.

\begin{figure}[bt]
\centerline{ \epsfxsize = 3.25truein \epsffile{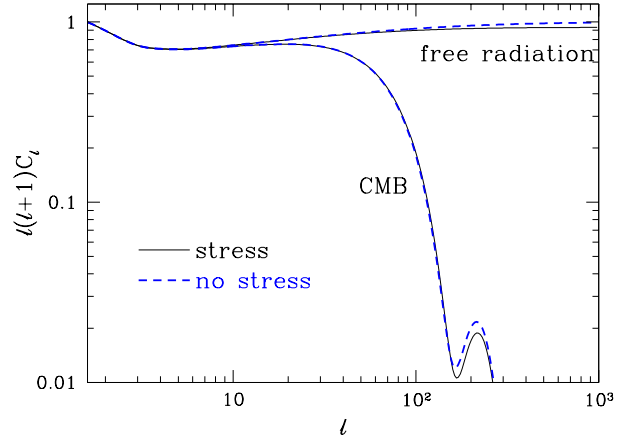}}
\caption{The effect of tensor anisotropic stress on tensor 
anisotropies.  Gravity waves generate tensor anisotropic stress
in radiation that absorbs energy and damps the gravity wave.
For free radiation like the neutrino background radiation,
this reduces the anisotropies generated on small scales.  For the CMB,
the elimination of tensor anisotropic stresses by scattering
cuts off small-scale contributions as well.
 }
\label{fig:cltensor}
\end{figure}

{\it Applications.---}
We show the damping effect of anisotropic stress in the
radiation backgrounds in 
Fig.~\ref{fig:cltensor}.  It is
generally not included in standard Boltzmann codes that solve
CMB anisotropies \cite{SelZal96}.  The reason is that 
it is negligible at large angles
since the corresponding modes entered the horizon well into
matter domination when the anisotropic stresses are
negligible. 
Since this feedback effect is typically small, equation~(\ref{eqn:tensorgreens})
may in principle be used to iterate to a solution from the free-gravity wave case of
equation~(\ref{eqn:thetaelltensor}).  However, the evolution of
the anisotropic stress has a more important effect.  
Just as scalar anisotropic stresses in the photons are destroyed by scattering
before last scattering (see \S \ref{sec:viscous}), 
tensor anisotropic stresses are destroyed as $e^{-\tau}$.  This cuts
off the tensor contributions to the anisotropies before last scattering as 
indicated in equation~(\ref{eqn:thetaelltensor}) and shown in 
Fig.~\ref{fig:cltensor}.
The feedback effect still exists, but the level of the anisotropy itself makes it
too low to be observable for reasonable tensor to scalar ratios.  

The Green's function solution of equation~(\ref{eqn:tensorgreens}) is more directly
applicable to the seed stress case where $p\Pi^{\pm 2}=\pseed\Piseed$.
As in the vector stress case, the phenomenological result is that defect models
tend to have significant tensor contributions above the angle the horizon 
subtends at last scattering.
As in the passive models, their contribution is cut off below this
scale due to scattering. 

\section{A Designer Application}
\label{sec:designer}

With this general study in hand, we are now in a position to
discuss prospects for reverse engineering the model
for structure formation.  
Obviously, the specific route the inversion takes will depend on
the results of ongoing experiments.  
Currently, the data favor a model with phenomenology 
like the $\Lambda$CDM model, e.g. the shape of 
the large-scale structure power spectrum
\cite{OstSte95,LidLytViaWhi96},
relative high to low redshift cluster abundances 
\cite{CarMorYeeEll97,BahFanCen97}, supernova luminosity distances
to redshift of $z \sim 0.5$ \cite{Rieetal98,Peretal98,Garetal98}, 
and degree scale CMB anisotropies (e.g. \cite{Lin98,Teg98}).
If agreement between the $\Lambda$CDM model and 
future precision tests is good, can we 
say purely on phenomenological grounds that we have proven the
existence of a cosmological constant?  If the model varies
from the data, can we use the methods developed here
to modify the stress history, restore agreement, and
in the process learn new information about the dark sector? 

\begin{figure}[bt]
\centerline{ \epsfxsize = 3.25truein \epsffile{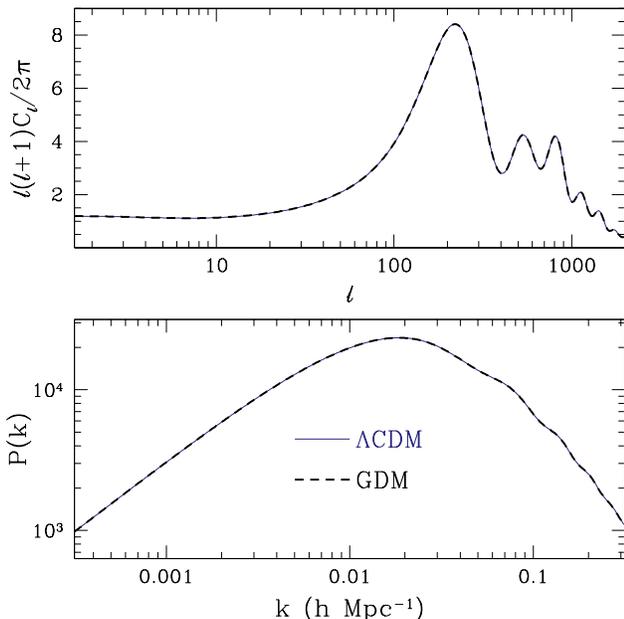}}
\caption{Mimicking $\Lambda$CDM.  Models with the same stress
histories in the dark sector have the same observable 
consequences regardless of differences in how that sector is
composed.  Here numerical solutions for the CMB anisotropy
and large-scale structure power spectra in a $\Lambda$CDM
universe and a single-component critical GDM dominated universe
of equation~(\protect{\ref{eqn:mimicconditions}})
are shown to be identical.  However, variations in $c_C$,
$\Delta \Phi$, and $\Delta \eta$ away from this tuned 
stress history do have observable consequences.}
\label{fig:mimic}
\end{figure}

Let us address the first question.  We know that the evolution
of structure is completely defined by the stress history of
the matter.  Since the equations of state of the ordinary
matter are known, the remaining element is the dark sector.  
To test the uniqueness of the 
$\Lambda$CDM model, we should look for alternate means of reproducing its
stress history.  
The background stress history of the dark sector in a $\Lambda$
model is given by
\begin{equation}
w_{\Lambda{\rm CDM}} = {-a^3 \over a^3 + 
	\Omega_{\rm CDM}/\Omega_{\Lambda}} \,,
\end{equation}
and its stress perturbations vanish.
This suggests that a generalized dark matter (GDM) component
of the type introduced by \cite{Hu98} 
and parameters
\begin{equation}
w_{\rm GDM} = w_{\Lambda {\rm CDM}}, \qquad c_{\rm GDM}^2 = 0  \,,
\label{eqn:mimicconditions}
\end{equation}
should reproduce the phenomenology of the $\Lambda$CDM model.
Recall that
$c_{\rm GDM}$ is the sound speed in the frame comoving with the
GDM (see \S \ref{sec:entropygen}).  To the extent that
the comoving and GDM-comoving 
frames coincide, this form of dark matter
exactly reproduces any mixture of $\Lambda$
and CDM.  We show an example in Figure \ref{fig:mimic}. 
Note that all classical cosmological and linear theory tests
will return the same answer for the two models despite the
fact that the GDM model is a single dark matter component model
in a critical density universe!  
Non-linear effects are also identical if the same stress history is
maintained throughout.

There are two lessons here.  The first is that on purely 
phenomenological grounds we can do no more than measure the global
properties of the dark sector.  A multicomponent model and 
a single component model with the same stress history are formally
identical. 

The second is that reproducing the phenomenology
of a $\Lambda$CDM model is rather simple: it requires an 
equation of state that varies from $w=0$ at high 
redshift to $w\approx -2/3$ today 
and a form of matter that is free of
large-scale stress gradients in its comoving frame.  
Thus, there exists a wide class
of such single component models that fit the current data as
well as the $\Lambda$CDM model. 

On the other hand, variations on the conditions in
equation~(\ref{eqn:mimicconditions}) have observable
consequences for future measurements. 
Relaxing the stress-free perturbation condition by raising
the sound speed reduces the small-scale power in the model 
and delays the formation of high redshift objects.
The remaining freedom in the stress history is
associated with the equation of state $w$.
We can quantify  
this by recalling that
the gravitational potential
depends only on the quantity
(see equation~[\ref{eqn:phicluster}]) 
\begin{equation}
\Delta \Phi \propto 
{\sqrt{\rhoc}\over a}\int_0^a{da\over\sqrt{\rhoc}} \,.
\label{eqn:phiintegral}
\end{equation}
Large-scale structure constrains the value of this
integral at the present ($a=1$).  CMB anisotropies from the ISW 
effect are sensitive to variations in this function 
that occur on the order
of the light-travel time across a wavelength (see \S 
\ref{sec:curvtoobs}).
Since CMB anisotropies potentially probe 
nearly three orders of magnitude from the current horizon,
variations at a fraction of a percent of the current expansion
time are potentially visible in the CMB. 

Similarly,
distance measures such as the angular diameter distance to 
the last scattering surface ($a \sim 10^{-3}$) 
and the luminosity distance
to high redshift objects ($a \sim 2/3$)
probe the combination 
\begin{equation}
\Delta\eta(a) \propto \int_a^1 da{ 1 \over a^2 \rho^{1/2}}.
\label{eqn:etaintegral}
\end{equation}

Current measures of $\Delta \Phi(a)$ and $\Delta\eta(a)$ are
crude at best as they only proble their values at discrete
epochs or averaged over long timescales.
A sharp test of the $\Lambda+$CDM hypothesis that should be
possible with future measures involves reconstructing the
time evolution of the equation of state $\wcr$ through 
measures of $\Delta\Phi(a)$ and/or $\Delta\eta(a)$. 
Any combination of a cosmological
constant and CDM will obey
\begin{equation}
\wcr' = 3\wcr(1+\wcr)\,.
\end{equation}  
The physical implication of this relation is that the background
pressure is constant in time $(p'=0)$ 
so that the adiabatic sound speed
vanishes $(c_s^2=p'/\rho'=0)$. 
If this condition is violated, we will have proven
the existence of a new form of matter.  
If $\wcr' > 3\wcr(1+\wcr)$, then it can be supported 
by adiabatic stresses \cite{ChiSugNak97}; 
if not, then a form of matter with non-adiabatic
stresses is required.  Non-adiabatic stresses imply that
the relation between the pressure and density perturbations does
not follow that of their spatial averages.

The simplicity of the requirements of 
equation (\ref{eqn:mimicconditions}) raises the possibility of a
unified description of the dark sector.
A concrete but somewhat trivial
example is 
a scalar field that rapidly oscillates around a non-zero
potential minimum.  The rapid oscillations average away all 
large-scale pressure effects save from the vacuum pressure 
of the potential minimum \cite{KhlMalZel85}. 
Unfortunately, the relationship between the constant and
quadratic pieces of the scalar potential is left unexplained
and so is no better than an explicit $\Lambda$+CDM model. 
Nonetheless, there may be more complicated examples,
perhaps involving multiple fields, in which the mimic 
conditions (\ref{eqn:mimicconditions}) are approximately
satisfied and that do unify the two behaviors in a true
sense.

This discussion shows that a reverse-engineered model for structure
formation will in general not be unique.  On the other hand, 
the observables can be translated into constraints on the 
stress histories and phenomenological models of the dark sector.
These in turn can assist in the search for compelling
physical candidates to compose the dark sector.  

\section{Discussion}
\label{sec:conclusions}

Without any assumptions besides the validity of general
relativity and nearly homogeneous and isotropic initial conditions,
the evolution of structure is completely determined by the
stress history of matter.  
We have studied the means by which stresses, both
in the background and the fluctuations,
can alter the observable properties of the model.

We have examined the effects of smooth, 
anisotropic, sonic, and entropic
stresses in structure formation, including their interactions
and ability to generate effectively smooth density components.  
We have illustrated these behaviors with analytic solutions
for systems with multiple components of differing
background equations of state, which can themselves be time dependent
in several important cases.   
These solutions have
applications to nearly all of the current models for structure
formation and are substantially more general than those existing
in the literature. 

Although this study is not exhaustive,
we have made explicit all of the assumption required
to arrive at specific models and their accompanying 
phenomenology.
In the process, we have exposed the limitations of
traditional categorization schemes
like that in Fig.~\ref{fig:taxonomy}.
These distinctions
can be blurred in cases where the usual assumptions 
do not apply.  We summarize several notable cases here:

\vskip 0.25truecm

\noindent{Initial Conditions:}
\vskip 0.25truecm

{\it Isocurvature initial conditions imply a growing comoving
curvature outside the horizon on scales relevant to large-scale
structure and degree-scale anisotropies.}  
The comoving curvature grows outside the horizon only 
by the action of stress perturbations.  Once stress perturbations are 
turned off, the curvature remains constant until horizon crossing 
or curvature domination.  
These considerations provide a means for mimicking the phenomenology
of adiabatic models \cite{Hu98b} and 
are important for interpreting the implications of 
CMB acoustic
peak phenomenology; however, they do not invalidate the conclusions
of \cite{HuWhi96}.

{\it The Newtonian curvature $\Phi$ 
is simply proportional to the
comoving curvature if the background equation of state is
constant.} The Newtonian curvature admits a decaying mode 
whereas the comoving curvature does not.  The decaying mode
can be stimulated by anisotropic stress perturbations outside
the horizon but has observable consequences only through the 
contribution remaining at horizon crossing \cite{Bar80}.
 
{\it If the Newtonian potential $\Psi$ is constant from
last scattering to the present, the observed temperature perturbation 
depends only on the equation of state and $\Psi$.}
The assumption here is that the comoving 
temperature perturbation is negligible and is only true if 
stress perturbations are also negligible compared with the 
comoving curvature for all time.  
The axion isocurvature
model provides a simple counterexample.
On the other hand, no assumptions about the
anisotropic stress are necessary even when 
$\Phi = -\Psi$ no longer holds. 

{\it Isocurvature initial conditions predict an observed temperature
perturbation of $2\Psi$ on scales larger than the horizon at last
scattering.}  The assumption here is that the initial temperature
perturbation in Newtonian gauge vanishes.  This is not the
case for models where the isocurvature conditions are established
by balancing the photon density perturbations off another species
of radiation.  Furthermore, changes in the potential that are
slowly varying compared with the light travel time across a
perturbation do not affect the observed temperature.

\vskip 0.25truecm
\noindent{Clustering Properties:}
\vskip 0.25truecm

{\it The smoothness of a component is gauge-dependent 
and hence has no physical meaning.}   The gauge dependence of
a smooth component is not a problem per se as certain frames,
e.g. the frame where the momentum of the component vanishs,
are dynamically special.  A smooth contribution to the
density with $w_S\ne -1$ 
does violate covariant energy conservation in any
coordinate system where the spatial curvature changes.  Since
the very presence of a smooth density component requires
that the comoving curvature perturbation decays, there can
be no identically smooth contributions in those coordinates except
in the trivial zero curvature perturbation case.  
A component can be smooth relative to another
species inside the horizon where the relativistic effects of
curvature variation are negligible \cite{QPaul}.

{\it The behavior of a smooth component depends only on its
equation of state $w_S$.}  Since all components 
except $\Lambda$ and curvature
are clustered outside the horizon, the manner
in which a component becomes smooth is observable.  
The presence of sonic (supportive) and anisotropic (dissipative)
stresses are two possibilities, but others exist in principle.

{\it A ``clustered'' model like standard CDM has no dynamically
important smooth stresses or density contributions.}
The radiation backgrounds and baryons are effectively smooth
well inside the horizon prior to recombination.
Smooth contributions are generic to structure formation models.

{\it Smooth component behavior implies small density perturbations
$\delta \rho_S < \delta \rho_C$.}
A component can be effectively
smooth even while possessing large density
fluctuations.  The crucial assumption is that their
time-average density over the dynamical time of the clustered
species is smooth.  Density perturbations in a component 
that vary rapidly with the expansion time generally lead to no
effect on the growth of structure.  The radiation backgrounds
mentioned above provide a familiar example. 

{\it The missing mass (clustered dark matter) and missing 
energy (smooth $w_S < -1/3$ dark matter) are separate problems.}
The stress history of the dark sector completely defines its
properties for classical cosmology and structure formation.  
Any combination of components that produces the same 
stress history will produce the same phenomenology.
As an example, we have constructed a toy model that
that exactly reproduces the $\Lambda$CDM phenomenology but employs
a single component of dark matter in a critical density universe. 
Variations in the stress history produce models that satisfy 
the current constraints equally
well but are potentially distinguishable from $\Lambda$CDM.

\vskip 0.25truecm
\noindent{Perturbation Type:}
\vskip 0.25truecm

{\it Scale-invariant gravity waves preferentially enhance 
the low-order multipoles of CMB anisotropies.}  Enhancement
only occurs if the gravity wave amplitude decays close to
horizon crossing and is eliminated as the equation of state
of the background drops.

{\it There is a sharp distinction between active and passive
models for structure formation.}  
Models that have been labeled ``passive''
in the literature are those in which the stress perturbations
are simply related to density, velocity, and metric fluctuations by 
equations of state, sound speeds, viscosity parameters, etc. 
Models in which there is a component whose stresses have no
fixed relation to density and metric perturbations have
been labeled ``active''.   
The issue is the number of internal degrees of freedom that 
act to specify the stresses.   For example, scaling
defect stress histories are typically approximated by tens of 
parameters and those of particle dark matter by three or fewer.
In principle, there is a spectrum of possibilities between
the two and a corresponding spectrum of phenomenological
consequences.

\vskip 0.5truecm

Our study is useful even if the current evidence supporting
$\Lambda$CDM-type phenomenology holds up.
Even the $\Lambda$CDM model itself does not have
an entirely trivial stress history as its dark sector 
includes the neutrino background radiation.  The observability of
the neutrino stress history has been addressed numerically 
in \cite{HuEisTegWhi98}; we have examined its physical origin
here.   Furthermore, the dark sector could contain exotic
features that produce more or less ordinary phenomenology,
and one needs to construct sharp tests against alternatives.
For example, a combination of cold dark matter and a cosmological
constant must obey
$\wcr' = 3\wcr(1+\wcr)$.  This relation also acts as the dividing
line between models with exotic and ordinary matter.
For exotic matter, the stress 
and density perturbations obey a relation that opposes that between
the background stress and density; scalar fields are one example
of exotic matter.

In summary, a purely phenomenological 
reverse-engineering of the model for structure formation will 
require the reconstruction of the time-averaged stress history of 
the dark sector.  This inversion is generally not unique.  
Nonetheless, if the observed phenomenology remains close
to that of our simplest models, our study of stress phenomenology 
should provide the means for constructing viable models.
If the observations require more radical
departures, our study should be useful in identifying 
the assumptions that are incorrect and assist in the search
for the correct phenomenological model for structure formation. 

\acknowledgements 
We would like to thank D.N. Spergel \& P.J. Steinhardt for
useful conversations.  W.H.\ is supported by the Keck 
Foundation and a Sloan Fellowship; D.J.E.\ by a Frank and 
Peggy Taplin Membership; D.J.E.\ and W.H.\ by NSF-9513835.

\end{document}